\documentclass[iop]{emulateapj}

\usepackage{natbib}
\usepackage{epsf}
\usepackage{color}
\usepackage{amsmath}
\newcommand{\spitzer}{\textit{Spitzer}}
\newcommand{\lir}{\hbox{L$_\mathrm{IR}$}}
\newcommand{\lpah}{\hbox{L$_\mathrm{PAH}$}}
\newcommand{\lsol}{\hbox{L$_\odot$}}
\newcommand{\lsim}{\mathrel{\hbox{\rlap{\lower.55ex \hbox{$\sim$}} \kern-.3em \raise.4ex \hbox{$<$}}}}
\newcommand{\gsim}{\mathrel{\hbox{\rlap{\lower.55ex \hbox{$\sim$}} \kern-.3em \raise.4ex \hbox{$>$}}}}

\begin{document}

\submitted{Accepted for publication in ApJ}

\title{Spitzer Spectroscopy of Infrared-Luminous Galaxies:  Diagnostics of AGN and Star Formation and Contribution to Total Infrared Luminosity}

\author{\sc Heath V. Shipley\altaffilmark{1}, 
Casey Papovich\altaffilmark{1},
George H. Rieke\altaffilmark{2}, 
Arjun Dey\altaffilmark{3},
Buell T. Jannuzi\altaffilmark{2,3},
John Moustakas\altaffilmark{4},
Benjamin Weiner\altaffilmark{2}}
\altaffiltext{1}{George P. and Cynthia Woods Mitchell Institute for Fundamental Physics and Astronomy, and Department of Physics \& Astronomy, Texas A \& M University, College Station, TX 77843-4242; heath.shipley@tamu.edu}
\altaffiltext{2}{Steward Observatory, University of Arizona}
\altaffiltext{3}{National Optical Astronomy Observatory, Tucson, AZ}
\altaffiltext{4}{Department of Physics \& Astronomy, Siena College, Loudonville, NY 12211}

\begin{abstract}

\noindent We use mid-infrared (MIR) spectroscopy  from the \spitzer\
Infrared Spectrograph (IRS) to study the nature of star-formation and
supermassive black hole accretion for a sample of 65 IR-luminous
galaxies at $0.02 < z < 0.6$ with $F(24\micron) > 1.2$~mJy.  The MIR spectra cover wavelengths
5-38$\micron$, spanning the polycyclic aromatic hydrocarbon (PAH)
features and important atomic diagnostic lines.  Our sample of
galaxies corresponds to a range of total IR luminosity, \lir\ =
L(8-1000\micron) = 10$^{10}$-10$^{12} L_\odot$ (median \lir\ of
3.0$\times 10^{11}$L$_{\odot}$).  We divide our sample into a
subsample of galaxies with \spitzer\ IRAC 3.6--8.0~\micron\ colors
indicative of warm dust heated by an AGN (IRAGN) and  those galaxies
whose colors indicate star-formation processes (non-IRAGN).  Compared
to the non-IRAGN, the IRAGN show smaller PAH emission equivalent
widths, which we attribute to an increase in mid-IR continuum from the
AGN.  We find that in both the IRAGN and star-forming samples, the
luminosity in the PAH features correlates strongly with [\ion{Ne}{2}]
$\lambda$12.8\micron\ emission line, from which we conclude that the
PAH luminosity directly traces the instantaneous star-formation rate
(SFR) in both the IRAGN and star-forming galaxies.  We compare the
ratio of PAH luminosity to the total IR luminosity, and we show that
for most IRAGN star-formation accounts for 10-50$\%$ of the total IR
luminosity.  We also find no measurable difference between the PAH
luminosity ratios of L$_{11.3}$/L$_{7.7}$ and L$_{6.2}$/L$_{7.7}$ for
the IRAGN and non-IRAGN, suggesting that AGN do not significantly
excite or destroy PAH molecules on galaxy-wide scales.  Interestingly,
a small subset of galaxies (8 of 65 galaxies) show a strong excess of
[\ion{O}{4}] $\lambda$25.9$\micron$ emission compared to their PAH
emission, which indicates the presence of heavily-obscured AGN,
including 3 galaxies that are not otherwise selected as IRAGN.  The
low PAH emission and  low [\ion{Ne}{2}] emission of the IRAGN and
[\ion{O}{4}]-excess objects  imply the  IR luminosity of these objects
is dominated by processes associated with the AGN.  Because these
galaxies lie in the ``green valley'' of the optical color-magnitude relation and have low implied SFRs, we argue their hosts have declining SFRs and these objects will transition to the red sequence unless some process restarts their star-formation.

\end{abstract}


\section{INTRODUCTION}
\label{intro}
The mid-IR (3-19\micron) spectrum of star-forming galaxies is dominated by strong emission features often associated with polycyclic aromatic hydrocarbons (PAHs).  PAH molecules are made up of tens to hundreds of carbon atoms in planar lattices that typically are several \AA\ in diameter \citep{Leger1984, Allamandola1985}.  Specifically, each PAH band is identified with specific carbon-carbon (C-C) and carbon-hydrogen (C-H) bending and stretching modes \citep{Allamandola1989}.  Their vibrational modes are excited by efficient absorption of UV and optical photons whose energy is re-emitted in the IR.  In many galaxies, the emission of young stars and active galactic nuclei (AGN) is almost entirely absorbed by dust and re-emitted in the IR; PAHs must then be used for detailed diagnostics of the luminosity sources.

The luminosity in the mid-IR PAH emission bands is very strong for galaxies with ongoing star-formation.  The total PAH emission can contribute as much as 20$\%$ of the total IR luminosity and the 7.7\micron\ PAH band may contribute as much as 50\% of the total PAH emission \citep[e.g.,][]{Smith2007}.  Certain PAH bands are made up of several dust emission features (e.g., 7.7\micron, 8.6\micron, 11.3\micron, 12.7\micron, and 17.0\micron\ PAH bands) and so we will use the general term ``feature'' to describe the PAH emission bands.

From laboratory experiments, models of stochastic heating of dust grains predict that the relative strengths of PAH features are dependent on the size distribution of PAH grains and the ionization state of the PAH molecules \citep{Schutte1993, DraineLi2001}.  Although the detailed physics is not well understood, existing models do explain the general trends observed for the various PAH features.

Previous \spitzer\ studies \citep{Diamond-Stanic2010, ODowd2009,
Smith2007} have found trends between the ratios of the various PAH
features and galaxy properties, such as AGN activity and star
formation history.  A tendency for active galaxies to have lower
ratios of the 6-8\micron\ PAH features relative to the 11.3\micron\
PAH feature has been reported both by \citet{ODowd2009} and
\citet{Smith2007}.  Smith et al.\ argued that AGN are able to modify the PAH
grain size distribution or are able to excite the emission of larger
PAH molecules.  O'Dowd et al.\ interpreted their results as being
consistent with the destruction of smaller PAH grains by shocks or
X-rays associated with the AGN, but were unable to determine a
difference between AGN-dominated and AGN--starforming composite
sources.  \citet{Diamond-Stanic2010} chose a sample of local Seyfert
galaxies to explore the effects of AGN on PAH emission.  They found
similar results to the previous studies when observing the
AGN-dominated nuclear regions of galaxies, and determined AGNs do not
excite PAH emission. In contrast, off-nuclear (presumably)
star-forming regions in the galaxy disks had 7.7-to-11.3\micron\ PAH
ratios similar to those of  star-forming galaxies in studies from the
literature.

Here, we extend studies of IR luminous galaxies with mid-IR
spectroscopy to higher redshifts and higher luminosities.  Specifically, we focus on
galaxies that  have high IR luminosities ($>10^{11}$L$_{\odot}$) and
moderate redshifts ($z \lsim 0.6$).  This allows us to study the
nature of the IR emission in galaxies at the same redshift range and
IR luminosity range that dominate the IR luminosity density at high
redshifts \citep[$z \sim 1$][]{LeFloc'h2005}.  By observing the mid-IR
spectral features of these galaxies we can therefore study the
ionization state of the IR emission, and study how both star-formation
and AGN contribute to the IR luminosity in these objects.  

An additional goal of our study is to understand if there is a
correlation (causal or otherwise) between the  IR-active stages
involving star-formation and/or AGN and other galaxy properties, such
as their optical color.  At a basic level the optical color depends on
the galaxies' star-formation histories \citep[galaxies with recent
star-formation have bluer optical colors, and passive galaxies lacking
any recent star-formation form a well documented ``red sequence'' e.g.,][]{Bell2004, Blanton2003, Faber2007}.  Studies of the optical colors of galaxies with high IR-luminosity (as measured by their MIPS  24\micron\
emission) show they span a wide range of optical color from the blue
star-forming galaxies,  through the ``green valley'', and up to the
blue edge of the red sequence \citep{Bell2005, Mendez2013}.  This highlights a
dependency of galaxy colors on IR activity. \citet{Wyder2007} show
that the sequence of star-forming galaxies becomes redder with
increasing optical luminosity, which is presumably a result of
increased dust obscuration.   This implies that stages of increased
dust obscuration in luminous galaxies may be a precursor to the
cessation of star-formation in galaxies as they migrate through the
``green valley'' to the quiescent red sequence \citep{Faber2007}, and
this may be related to quenching by the presence of an AGN \citep{Bundy2008, Nandra2007, Schawinski2007, Schawinski2010}.  Because IR-luminous galaxies have optical colors that
span from the star-forming sequence to the red-sequence,  there may be
a relationship between their excitation mechanism (star-formation
versus black hole accretion) and their optical color \citep{Bell2005, Chen2010, Nandra2007}.  

To date samples of IR-luminous with mid-IR spectroscopy have been
heterogeneous or incomplete in their coverage of galaxies over the full
range of optical color.  For our study, we allow for changes in the mid-IR spectral features as a function of optical color
in galaxies by using a sample of IR-luminous galaxies that is selected
to have a uniform distribution in rest-frame $(u-r)_{0.1}$\footnote{Note that throughout, $(u-r)_{0.1}$ and $M(r)_{0.1}$ are the rest-frame color and absolute magnitude observed at z=0.1 \citep[as used extensively in the SDSS literature, see, e.g.,][]{Blanton2003}.  Furthermore, throughout, $M(r)_{0.1}$ and $(u-r)_{0.1}$ are measured in AB magnitudes \citep{Oke1983}.} color.  In sections \ref{data} and \ref{spitzer imaging}, we describe our sample selection and ancillary data.  In
section \ref{analysis}, we describe our \spitzer\ IRS observations,
data reduction, and analysis of the IRS spectra.  In section
\ref{Results}, we describe our measurements of the total IR luminosity
and discuss correlations between the PAH emission and emission from
other atomic lines. In section \ref{discussion}, we discuss the
color-magnitude relation for IRAGN, AGN effects of PAH emission, and
the significance of [\ion{O}{4}] emission.  In section
\ref{conclusions}, we present our conclusions.  The $\Lambda$CDM
cosmology we assume is H$_{\mathrm{0}}$ = 71 km s$^{\mathrm{-1}}$
Mpc$^{\mathrm{-1}}$, $\Omega_m$ = 0.27, and $\Omega_\lambda$ = 0.73
throughout this work.

\section{Sample Definition and Selection}
\label{data}
\label{sample selection}
Our main goal is to compare the mid-IR spectral properties of AGN and star-forming (SF) galaxies in luminous IR galaxies (LIRGs) out to moderate redshifts ($z \lsim 0.6$).   To accomplish this goal, we must be sensitive to the fact that both AGN and IR-luminous star-forming galaxies are known to span a broad range in optical color (see section \ref{intro}).  As discussed above, the energetics of both obscured AGN and star-forming galaxies may affect both the mid-IR spectral properties and the optical color.  To avoid any bias related to these effects, we therefore built a sample of IR-luminous galaxies for spectroscopy that span an approximately uniform distribution of $(u-r)_{0.1}$ optical color that spans the full range from the blue, star-forming sequence to the red, quiescent sequence. 

We built our primary sample for IRS spectroscopy using the AGN and Galaxy Evolution Survey \citep[AGES,][]{Kochanek2012}.   This survey provides deep ($I<20$ mag) spectroscopy covering the $\sim$10 deg$^2$ Bo\"{o}tes field, which includes deep optical imaging from the NOAO Wide Deep-Field Survey (NWDFS) \citep{Jannuzi1999}.  The field was also covered with \spitzer/IRAC imaging at 3.6--8.0~\micron\ as discussed in \citet{Ashby2009}, and with MIPS imaging at 24--160~\micron\ first with shallow coverage by \citet{Houck2005} and subsequently with much deeper coverage as part of  MIPS AGES (MAGES) survey (PI: B.\ Januzzi).   The spectroscopy, optical, and mid-IR imaging allows us to select IR-luminous galaxies out to moderate redshift and over the full range of $(u-r)_{0.1}$ optical color.

As our parent sample, we selected 498 galaxies from the AGES data with spectroscopic redshifts 0.2 $<$ z $<$ 0.6,  and $f_{\nu}$(24\micron) $>$ 1.2 mJy.  We then searched the \spitzer\ archive for programs that had previously targeted galaxies in our parent sample with IRS spectroscopy.  Two programs satisfied our needs and included \spitzer/IRS data for a total of 17 galaxies from PID 20113 (PI: H. Dole) and 38 galaxies from PID 20128 (PI: G. Lagache); to our knowledge the data from these programs have not been published previously.

\begin{figure}
\epsscale{1.2}
\plotone{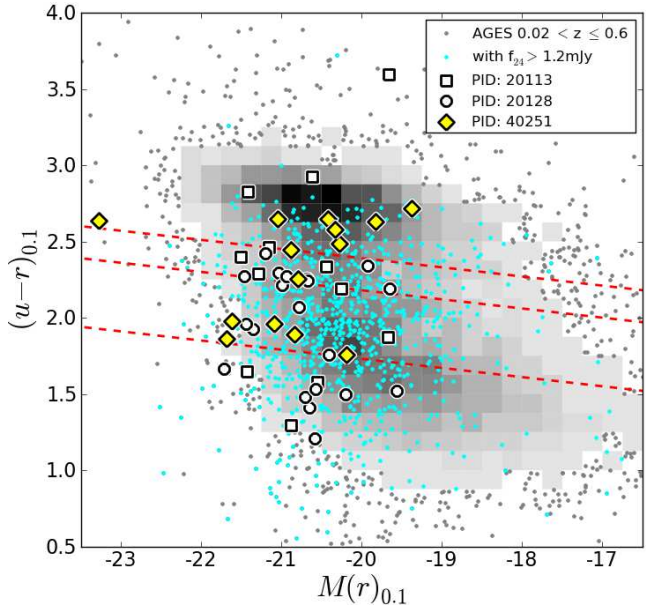}
\caption{Color-Magnitude Diagram using optical data from the AGES catalog for our sample.  The grey-shading indicates the density of all galaxies from AGES in that region of color-magnitude space and $0.02 < z < 0.6$, where the grey-shading increases as the density increases linearly.  The cyan points show those AGES sources with $0.02 < z < 0.6$ and $f(24\micron) \geq 1.2$~mJy, the IRS spectroscopic limit for our sample.  The galaxies selected in our sample are indicated by white squares, white circles, and yellow diamonds for Dole (program 20113), Lagache (program 20128, non-FLS sources), and Rieke (program 40251), respectively.  Furthermore, we selected IR-luminous galaxies for our sample such that they span the full range of $(u-r)_{0.1}$ optical color with an equal number (12-13) galaxies in each of four bins, denoted by the red-dashed lines and defined in section \ref{sample selection}.}
\label{CMD}
\end{figure}

We supplemented the archival IRS spectroscopy with new IRS observations of additional galaxies taken as part of the MIPS guaranteed time observations (GTO) 40251 (PI: G.~Rieke) to ensure that our full IRS sample includes galaxies with an approximately uniform distribution in $(u-r)_{0.1}$ color.  To identify the sample for GTO observations, we derived bins in $(u-r)_{0.1}$ optical color that run roughly parallel to the red sequence, which has some slope in the color-magnitude plane:  
\begin{align*}
(u-r)_{0.1} + 3/50 \times (M[r]_{0.1} + 20) > 2.39 \\
2.39 \geq (u-r)_{0.1} + 3/50 \times (M[r]_{0.1} + 20) > 2.18 \\
2.18 \geq (u-r)_{0.1} + 3/50 \times (M[r]_{0.1} + 20) > 1.73 \\
(u-r)_{0.1} + 3/50 \times (M[r]_{0.1} + 20) \leq 1.73,
\end{align*}
We require at least 12 galaxies in each of these color bins.   As shown in figure \ref{CMD}, this required new IRS observations of 14 additional galaxies from our GTO allocation in PID 40251.

The combined archival IRS data (7 + 19 galaxies)
and new IRS data (14 galaxies) yield a sample of 40 galaxies that form
our primary IRS sample.  In addition, the archival programs used here
include 29 additional galaxies, either at lower redshift (0.02 $< z <$ 0.2; 12 galaxies) or from the \spitzer\ First Look Survey (FLS; 17 galaxies), or both.  These galaxies appear to all be selected using a 24~\micron\ flux limit
($f_\nu(24\micron) > 1.2$~mJy).  We included these galaxies as a
secondary sample here because our sample is increased by $\sim$70\% and it
enables us to extend our observations to both lower-luminosity objects (in the case of the objects at 0.02 $< z <$ 0.2) and adds higher luminosity objects from the FLS (even though these have no
$(u-r)_{0.1}$ color information).  As discussed in section \ref{data
reduction}, we ultimately exclude four observations (two from AGES and
two from the FLS) that showed contamination from a serendipitous
source in the IRS slit.   Therefore, our full sample of primary and
secondary sources includes 65 galaxies with $f_\nu(24\micron) >
1.2$~mJy and $0.02 < z < 0.6$.

\begin{figure}
\epsscale{1.2}
\plotone{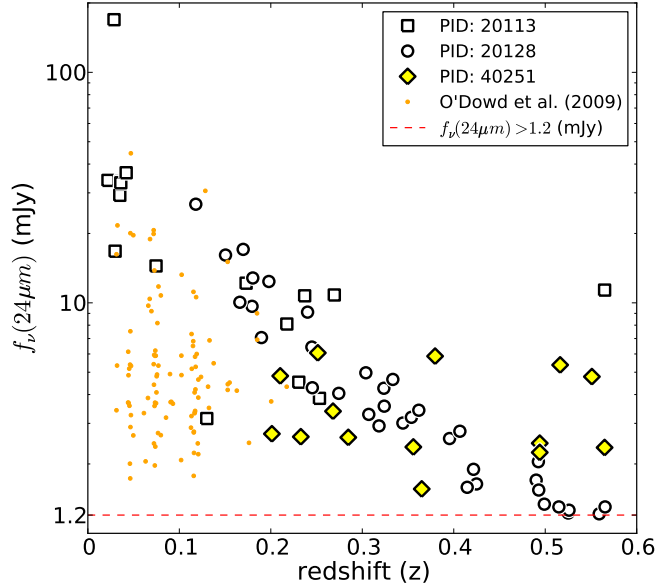}
\caption{Distribution of the redshifts and 24\micron\ flux densities of the 65 galaxies in our IRS sample.  The galaxies are indicated by white squares, white circles, and yellow diamonds for Dole (program 20113), Lagache (program 20128), and Rieke (program 40251), respectively.  The orange circles are from \citet{ODowd2009} sample as a comparison to our sample.  The red dashed line indicates $f_\nu(24\micron) = 1.2$mJy.  The redshift mean and median of the distribution are 0.30 and 0.28, respectively.   The redshift distribution is fairly uniform from $0.2 < z < 0.6$, with an interquartile range (which contains the inner 50\% of galaxies) of $z_\mathrm{interquartile} = 0.18 - 0.42$.}
\label{hist z}
\end{figure}

In summary, we have three samples defined in this paper.
\begin{itemize}

\item Full Sample.  All 956 galaxies with $f_\nu(24\micron) > 1.2$~mJy and redshift of 0.02 $< z <$ 0.6 in AGES.

\item Parent Sample.  All 498 galaxies with $f_\nu(24\micron) > 1.2$~mJy and redshift 0.2 $< z <$ 0.6, from which targets for new IRS observations were selected.

\item IRS Sample.  The 65 galaxies with $f_\nu(24\micron) > 1.2$~mJy that have IRS spectroscopy either from archival data (51 galaxies) or our new GTO observations (14 galaxies).

\end{itemize}

Figure \ref{CMD} shows the $(u-r)_{0.1}$--$M(r)_{0.1}$ color-magnitude diagram for all the galaxies in AGES in the redshift range of our full sample, and indicates the galaxies in the redshift range with ($f_\nu(24\micron) > 1.2$~mJy (956 galaxies).  The figure shows that the existing galaxies from the \spitzer\ archival programs (non-FLS galaxies) are biased toward IR-luminous galaxies with blue colors compared to the AGES sample and the sample of GTO observations help to uniformly distribute the galaxies in $(u-r)_{0.1}$ color.  The properties of our sample are listed in table \ref{table galaxy stats} and figure \ref{hist z} shows the redshift and 24\micron\ flux density distributions.

\section{\spitzer\ Imaging Data}
\label{spitzer imaging}

\subsection{Bo\"{o}tes Field}
\label{AGES data}

For our IRS sample, we used observations taken from the Infrared Array Camera (IRAC) and MIPS.  IRAC observations were taken from the \spitzer\ Deep Wide-Field Survey (SDWFS) catalog for Bo\"{o}tes sources \citep{Ashby2009}.  The sources in our sample had IRAC coverage in all four photometric IRAC bands (3.6$\micron$, 4.5$\micron$, 5.8$\micron$, and 8.0$\micron$).  We used the 8.0$\micron$ selected SDWFS catalog for the four IRAC band magnitudes (given in the Vega magnitude system) of the Bo\"{o}tes sources.\footnote{ Throughout we use [3.6], [4.5], [5.8], and [8.0] to denote the magnitudes with respect to Vega measured in each IRAC band, respectively.}

Since our IRS data were acquired, deeper MIPS data were taken as part of the MIPS AGES (MAGES) program (B. Jannuzi et al., in prep).  These data catalogs provide deeper imaging in all three MIPS bands, (24~\micron, 70~\micron, and 160~\micron) in the Bo\"{o}tes field.  The MAGES catalogs provide significantly deeper imaging at 70~\micron\ and 160~\micron, achieving detections at these wavelengths for many of the sources in our IRS sample.

Using the MAGES data catalog, we first matched all 50 galaxies (in the AGES Bo\"{o}tes' field) with a $r=2\arcsec$ matching radius at 24~\micron.  For all galaxies, the IRS integrated 24~\micron\ flux densities (see section \ref{data reduction}) agreed within the measurement errors with those matched to the  MAGES 24~\micron\ flux densities.   We then used the MAGES matched catalog to obtain 70~\micron\ and 160~\micron\ flux densities for sources in our IRS sample, including 46 galaxies with 70~\micron\ detections and 27 galaxies with 160~\micron\ detections (flux densities given in table \ref{table galaxy stats}).  We use these longer wavelength data to improve our estimates of the total IR luminosity (see section \ref{LIR}).

\subsection{\spitzer\ First Look Survey (FLS)}
\label{FLS data}
Sources in our IRS sample from the \spitzer/FLS have IRAC and MIPS imaging from programs published in the literature.  We used data from the four IRAC channels from the \spitzer\ data archive provided from \citet{Lacy2005}.  We used the combined FLS catalog to acquire the four IRAC band flux densities of the FLS sources.  We converted the flux densities to the appropriate magnitudes (Vega magnitude system), identical to that for the Bo\"{o}tes sources.  We use the catalogs of \citet{Papovich2006} and \citet{Frayer2006} to obtain MIPS 24\micron\ flux densities, and MIPS 70\micron\ and 160\micron\ flux densities, respectively.

\subsection{IRAC AGN Selection}
\label{AGN selection}
\begin{figure}
\epsscale{1.2}
\plotone{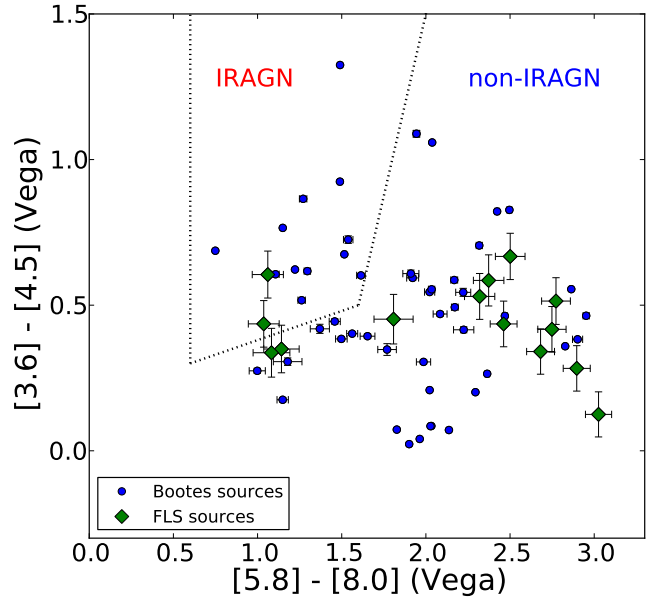}
\caption{IRAC colors of [5.8] - [8.0] versus [3.6] - [4.5] for galaxies in our IRS sample.    Here, the IRAC colors are in the Vega magnitude system, following Stern et al.\ (2005).  The blue circles are sources from the Bo\"{o}tes field and green diamonds are FLS sources.  The dotted lines show the empirical selection-criteria for IRAC-AGN selection: galaxies inside this ``wedge'' are IRAC-selected AGN \citep{Stern2005}.  We denote the subsample of galaxies in our IRS sample that satisfy these colors as IRAGN.  We denote galaxies outside this wedge as non-IRAGN, and we expect the IR emission in these objects to be dominated by star-formation.} \label{IRAC AGN}
\end{figure}

The primary goal of this study is to compare the mid-IR spectral features of IR-luminous galaxies with indications of AGN activity to those without.   IR-selected AGN can be identified from ``star-forming'' IR galaxies because AGN typically show indications of warm ($\sim$1000 K) dust heated by the AGN which produces a characteristic power law continuum through the IRAC channels \citep[e.g.,][]{Donley2012}.  We take advantage of this color selection here to identify IR-selected AGN using the color selection criteria proposed by \citet{Stern2005}.   The Stern et al.\ color selection is defined using spectroscopic observations of galaxies in the same redshift range as our sample, which makes it appropriate here.  This selection is also relatively immune to mis-identification of purely star forming galaxies as AGN over the redshift range of our study \citep{Donley2008}, so this region in the diagram can be re-interpreted to identify galaxies where the AGN makes a substantial contribution to the mid-IR output but is not totally dominant \citep{Mendez2013}.  We show the result of this selection on our IRS spectroscopic sample in figure \ref{IRAC AGN}. 

We define the subsample of galaxies in our IRS sample that satisfy this IRAC color-color selection as IRAGN.   This includes 14 out of the 65 galaxies in our IRS sample (2 are in the FLS sample and 12 are in the AGES Bo\"{o}tes sample).   Based on the statistics from \citet{Stern2005}, it is unlikely that our IRAGN subsample contains any star-forming galaxies for the AGES sample.  It is likely, however, that  our subsample of IRAGN is incomplete for all AGN in our sample.  Furthermore, the number of AGN dominated galaxies in our sample may be biased towards a higher fraction because of our 24\micron\ selection criteria.  For the remainder of this paper we refer to galaxies in our sample satisfying this selection as IRAGN, and the other galaxies in our sample as non-IRAGN.  We expect the non-IRAGN subsample to be dominated by star-forming galaxies even though some will have an AGN.

\newpage

\section{IRS Observations, Data Reduction, and Analysis}
\label{analysis}

\subsection{IRS Observations}
\label{obs}
The IRS was used in staring mode to make the observations with both the low-resolution short-low (SL) and long-low (LL) IRS modules, giving 5.2-38.0$\micron$ coverage (SL $\sim$ 5.2-14$\micron$ and LL $\sim$ 14-38$\micron$) with resolving power of $\sim 60-125$.  Exposure times were estimated based on the 24 $\micron$ flux densities and the purpose of the individual programs.  Observing times ranged from 360-600 seconds for the SL observations and 180-3600 seconds for the LL observations.  The SL and LL modules have respective slit widths of $\sim$3.7'' and $\sim$10.5'', the slits have physical sizes of 1.5-25 kpc and 4.3-70 kpc respectively, over the redshift range of the sample ($0.02 < z < 0.6$).  At the median redshift, the physical slit sizes are ~16 kpc and ~45 kpc, respectively.

\subsection{Data Reduction}
\label{data reduction}
We used the two-dimensional \spitzer\ data products processed by the \spitzer\ Pipeline version S18.7.0 to  perform the standard IRS calibration. Our  post-pipeline reduction of the spectral data started from the pipeline products basic calibrated data (bcd) files.  To perform sky subtraction, we created a sky frame for each object by taking the median of all the images for a single object at the same nod position.  We subtracted this image from the frames for the other nod position for the object to produce sky subtracted images.  We combined all sky subtracted images for each object at a given nod position to produce coadded 2-dimensional images.   We used IRSCLEAN (v.1.9) on each image to remove and correct for bad and rogue pixels.

We used the \spitzer\ IRS Custom Extractor (SPICE) software to extract one-dimensional spectra for each order at each nod position.  We chose to use the optimal extraction with point-source calibration because it improved the signal-to-noise (S/N) ratios for our sources.  We combined the 1D spectra manually using a weighted mean for the nod position for each order.   We combined the SL2 + SL1 orders (for the 25 sources with SL2 data) and LL2 + LL1 orders (for all sources).

To combine the SL and LL orders we took care to match the flux calibration, accounting for light lost outside the spectroscopic slits (which varies considerably between SL and LL, see above).    We integrated the IRS spectra from the SL module with the IRAC 5.8$\micron$ and the IRAC 8.0$\micron$ transmission functions, and we integrated the spectra from the LL module with the MIPS 24$\micron$ transmission function.  We took the ratio of the observed flux density as measured directly from the IRAC and MIPS observations to the flux density synthesized from the IRS spectra as an aperture correction (although other effects may contribute to variations in the IRS flux density) for each spectrum, with mean corrections of 1.55, 0.98, and 1.12 for IRAC 5.8\micron, IRAC 8.0\micron, and MIPS 24\micron\ respectively.  We combined (using a weighted mean, weighting by inverse variance) the flux-corrected spectra to produce a single 1D spectrum for each source covering the entire wavelength range.  

During the process of the IRS data reduction, we rejected four galaxies from our sample because these objects show strong contamination from another source in the slit, resulting in a dubious spectrum.  After rejecting these galaxies, our IRS sample includes IRS spectra of 67 objects used in the spectral analysis. However, two of our sources were repeat visits of the same source  (sources observed both by the 20113 Dole and 20128 Lagache programs).  Comparing the reduced spectra for each of these sources, they give consistent results for redshift and PAH ratios.  For the analysis here, we rejected the visits with lower S/N.  Therefore, our final dataset includes IRS spectra for 65 unique objects.

\subsection{IRS Spectral Fitting}
\label{fits}
To study the mid-IR emission features in the IRS spectra of our sample, we used the PAHFIT spectral decomposition code \citep{Smith2007}, designed for \spitzer\ IRS data.  PAHFIT uses a $\chi^2$ minimization routine to fit a non-negative combination of multiple emission features and continua to the one-dimensional spectra of our sources.  The features included in PAHFIT are the dust emission features from PAHs (modeled as Drude profiles), thermal dust continuum, continuum from starlight, atomic and molecular emission lines (modeled as Gaussians), and dust extinction.   The PAH emission features at (e.g. 7.7, 11.3, and 17 $\micron$) are blends of multiple components, and PAHFIT treats these complexes as individual emission ``features''.

\begin{figure}
\epsscale{2.4}
\plottwo{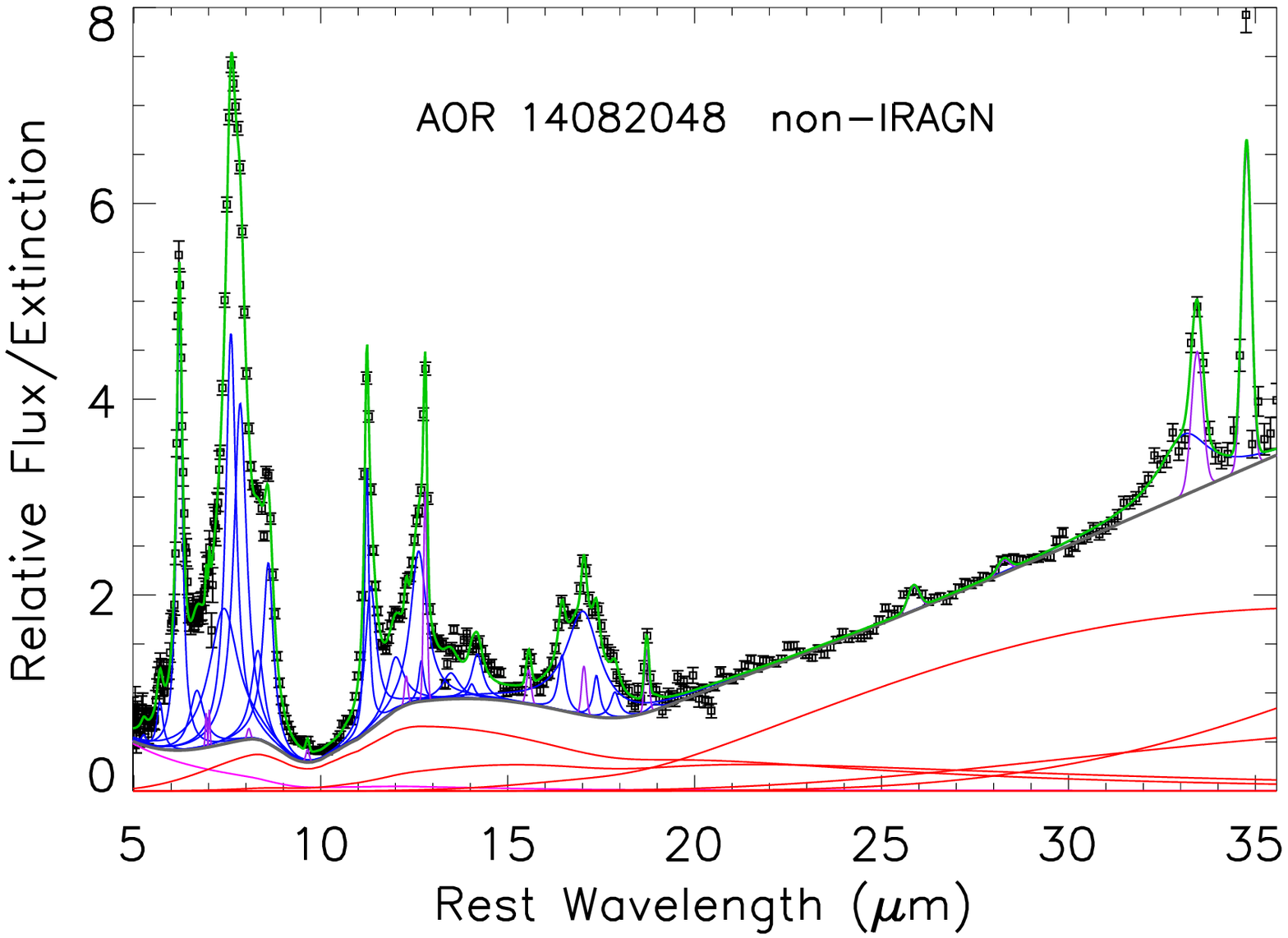}{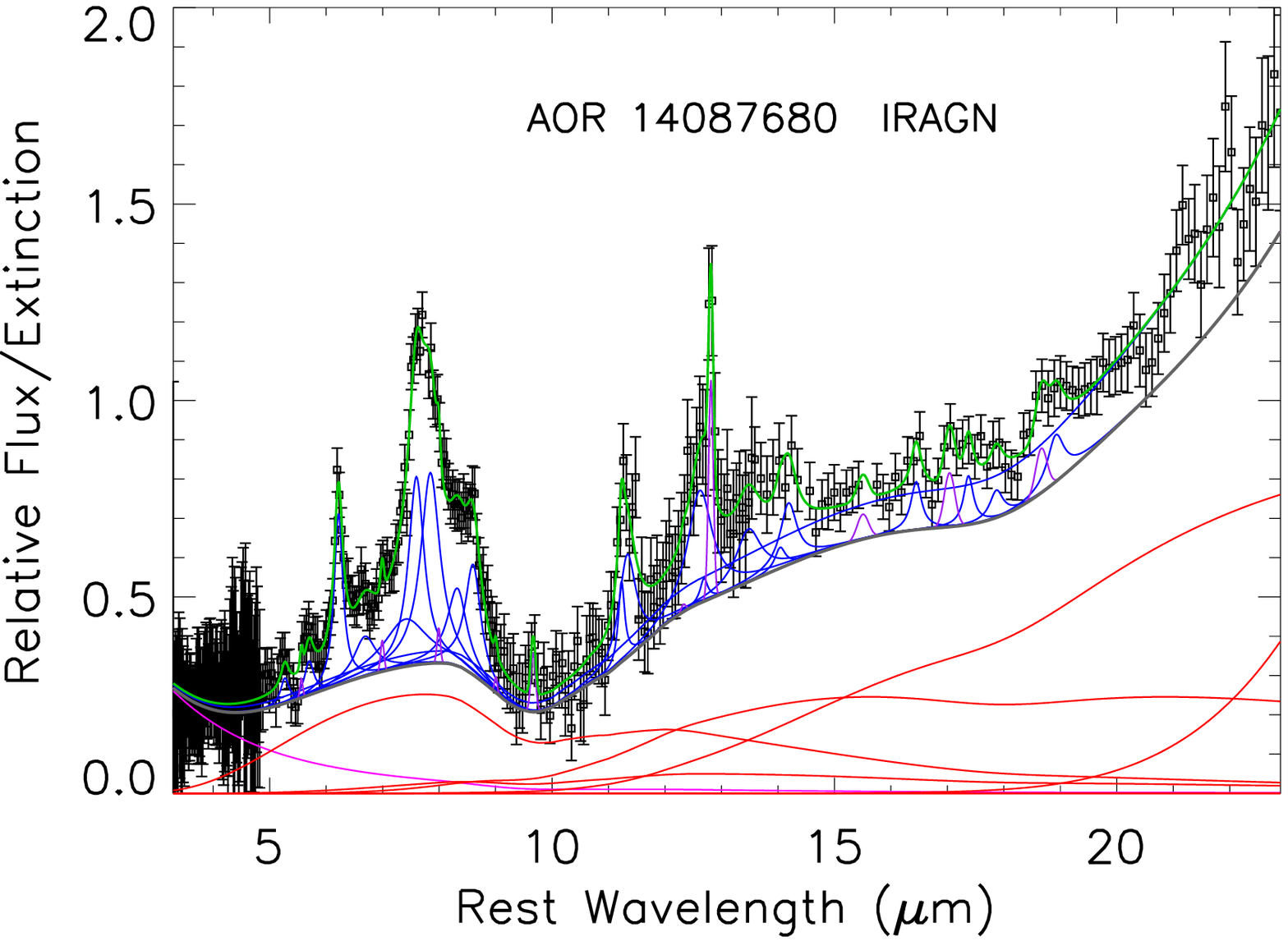}
\caption{Examples of PAHFIT spectral decomposition to object 23 (top) and object 4 (bottom).  In each panel, the IRS spectroscopic data are shown as black squares.   The total fit and individual spectral components fit by PAHFIT are shown, including the molecular and atomic emission features (blue curves), dust continua (red), and stellar light (magenta).  The total fit (the sum of all the model components) is shown in green, which provides a good representation of the data.}
\label{decomps}
\end{figure}

We used PAHFIT to fit the IRS spectrum for each galaxy in our sample.
The line fluxes derived by PAHFIT for the emission features for each
object are listed in Tables \ref{table PAH emission fits} and \ref{table line emission fits}.  Figure
\ref{decomps} shows the PAHFIT spectral decompositions for the spectra
of objects ID=4 and 23 in our sample, using the IDs in Table~\ref{table
galaxy stats}.  Object 4 is classified as an IRAGN as described by
the method in Section \ref{AGN selection}, while object 23 shows no
indication of an AGN.  The spectra are characteristic of the data quality and fit quality of our sample.

We fit the IRS spectra for the galaxies in our sample using PAHFIT
models with and without dust extinction.  PAHFIT uses a model where
the dust is mixed with the emitting stars and grains, as opposed to an
absorbing screen.   Our fits showed that the PAHFIT $\chi^2$ values
are improved using models with dust extinction for 60/65 galaxies
in our sample (the other 5 galaxies had similar $\chi^2$ values and are low S/N sources).  Other studies have also found it desirable to use the default PAHFIT dust extinction parameter \citep{Smith2007, ODowd2009, Wu2010,
Diamond-Stanic2010}.  Therefore, we chose to use results from PAHFIT
for our sample that include dust extinction.   However, none of our conclusions are changed if we use the fits excluding dust extinction.  

Using PAHFIT, we also fit for the redshift of each galaxy using the
IRS spectra.  All galaxies in our sample have a measured redshift from
optical spectroscopy from AGES in the Bo\"otes sample \citep{Kochanek2012} or from Papovich et al.\ (2006) in the FLS sample, but the PAHFIT measurement provides an independent check on the redshift, and
provides supporting evidence that the mid-IR source corresponds to the
optical counterpart.    PAHFIT requires an input redshift (it does not
minimize over redshift as part of the fit).   Therefore, for each
galaxy in our sample we used PAHFIT with a grid of redshift.  We did
this step iteratively.  We first ran PAHFIT using a the full redshift
range with $0.0 < z < 0.6$ with steps of $\Delta z = 0.01$, and we
selected the fit with the best reduced $\chi^2$.  We then refined the redshift solution by fitting the redshift with
$\Delta z = 0.001$ using a redshift range $\pm 0.1$ about the best-fit
redshift from the first step.  

We verified the consistency in our redshifts by doing a comparison to the AGES optical redshifts.  We found the two redshifts for all galaxies agree within 2\%, the resolution of the IRS.  Table~\ref{table galaxy stats} gives both the redshift measured from the optical spectroscopy and the redshift measured independently from the IRS spectrum here.

\subsection{Offset Between IRS SL and LL Modules}
\label{pah offset}

In the analysis below, one of the important emission-line ratios we will consider is the relative strength of the 7.7~\micron\ to 11.3~\micron\ PAH features.   For galaxies with $z < 0.2$, these lines both sit in the IRS SL module, but for $z \ge 0.2$, the 11.3~\micron\ feature shifts into the LL module, in which case the two lines are measured from different modules with different slit widths.   Our procedure to flux-calibrate the spectra using the IRAC and MIPS photometry described above should account for changes in the relative flux owing to the slit widths.  Nevertheless, we did some additional tests to confirm that the flux measured for the 11.3 \micron\ feature is accurate.

The median redshift of our IRS sample is $z_\mathrm{median} = 0.28$, and therefore, nearly 50\% of the galaxies in our IRS sample have the 11.3~\micron\ feature in the LL module.   For approximately one quarter of the sample (0.19 $<$ z $<$ 0.28), the 11.3~\micron\ feature lies in the overlap region covered by both the SL and LL modules.  For these galaxies, we compared the measured emission of this feature in the SL and LL modules separately, and found that for 13/16 galaxies there was less than a 3\% difference in the measured flux of the feature.  For the remaining three galaxies, the features have lower signal-to-noise, and the difference in the fluxes is within the uncertainties on the measurement.  Therefore, we are confident that the uncertainties are fully represented in the errors on the data, and that any systematic errors from our flux calibration from using the two modules are small. 

\section{Comparison of Mid-IR Emission Features and Relation to Total Infrared Luminosity}
\label{Results}

\subsection{Composite Spectra}
\label{composite}

\begin{figure}
\epsscale{1.2}
\plotone{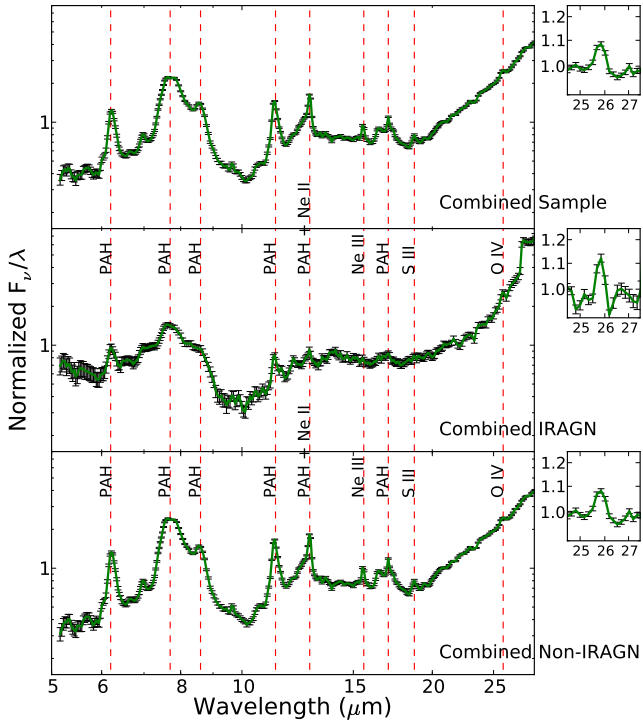}
\caption{Composite spectra for the IRS sample.  The top panel shows
the composite spectrum for all 65 galaxies in the IRS sample.  The middle
and bottom panels show composite spectra for the subsamples of IRAGN
(14 galaxies) and non-IRAGN (51 galaxies), respectively.  The vertical red-dashed
lines indicate the prominent PAH features and emission lines in the
wavelength range, as labeled.  The flux is normalized at the continuum flux of
21$\micron$.  The outset panels show the composite spectra in a small
wavelength region 24.5$\micron$-27.5$\micron$ to show the strength of the
[O IV] $\lambda$25.9$\micron$ emission line (see section \ref{composite} for explanation).  The error bars shown is the error on the weighted mean for each composite spectrum.}
\label{composite spec}
\end{figure}

The spectra of the IRAGN and non-IRAGN
subsamples show different continua and strength of emission features,
as illustrated in Figure \ref{composite spec}.   The figure shows
composite spectra for the full sample (IRAGN + non-IRAGN), the IRAGN, and non-IRAGN.  The composite spectra were created by performing a weighted mean.  First, the spectra were interpolated to a common wavelength spacing using the observed wavelength scale of each order for the SL and LL modules ($\Delta \lambda$(SL2) $\approx$ 0.030\micron, $\Delta \lambda$(SL1) $\approx$ 0.059\micron, $\Delta \lambda$(LL2) $\approx$ 0.083\micron, $\Delta \lambda$(LL1) $\approx$ 0.166\micron) to preserve the information from each spectrum when creating the composite spectrum.  The composite spectra are then normalized to the continua fluxes at 21\micron.  An outset panel of each composite spectrum is shown for the [\ion{O}{4}] $\lambda$25.9\micron\ emission line.  For the [\ion{O}{4}] panel, we have subtracted the continuum out (by fitting a line to the continuum around the [\ion{O}{4}] line for each composite spectrum) to better show the strength of the [\ion{O}{4}] line for the IRAGN compared to the non-IRAGN.  The [\ion{O}{4}] atomic line at 25.89\micron\ has been established to be an indicator of AGN luminosity by \citet{Melendez2008} and \citet{Rigby2009} by comparison to hard (E $>$ 10 keV) X-rays.  \citet{Diamond-Stanic2009} showed that this line can be used to find AGN that are in heavily dust-obscured galaxies and therefore are missed by optical, X-ray, and other selection criteria.

The IRAGN composite spectrum shows a lower PAH emission equivalent width compared to the
non-IRAGN composite spectrum.  The most apparent PAH emission feature
in the IRAGN subsample is the weak 7.7 $\micron$ feature.  The
continuum is mostly flat across the wavelength range of the spectra,
suggesting the continuum is raised in response to heating from the
AGN.  Furthermore, the PAH features are not absent in the IRAGN, which suggests
that in part, the mid-IR continuum from the AGN is diluting the
luminosity from the PAH features.  In comparison, the continuum is
much weaker in the non-IRAGN composite spectrum, and this spectrum
shows very prominent PAH and other emission features throughout the
wavelength range covered. 

\subsection{Measuring the Total Infrared Luminosity}
\label{LIR}

We used model spectral energy distributions (SEDs) to estimate the
total L$_{\mathrm{IR}}$ = L$_{\mathrm{8-1000} \micron}$ from the MIPS
24$\micron$ flux densities for all sources.  70\micron\ and 160\micron\ flux densities were used when available from MAGES for Bo\"{o}tes sources and \citet{Frayer2006} for the FLS sources.  To calculate the total IR luminosity, we tested several model libraries that provide far-IR SEDs that vary in shape as a function of galaxy IR luminosity and
ionization \citep{DaleHelou2002, CharyElbaz2001, Rieke2009}.   
For galaxies with 24 and 70~\micron\ detections, or 24, 70, and 160~\micron\ flux
densities, the total IR luminosities are consistent (within 40$\%$) using any of the
IR SEDs from  \citet{CharyElbaz2001}, \citet{DaleHelou2002}, or
\citet{Rieke2009}.  However, because 19 galaxies (including four IRAGN) in our sample are detected at only
24~\micron, we use the \citet{Rieke2009} IR SEDs because our tests
showed that the total IR luminosities derived using their templates with only the
24~\micron\ are closest with those derived from multiple MIPS bands
(24 and 70~\micron, or 24, 70, and 160~\micron).  These tests are described more fully in Appendix \ref{MAGES LIR}.  For the four IRAGN only detected at 24\micron, we interpret the luminosities from the purely star-forming spectral templates to be upper limits.  However, if we assume an extreme case of a quasar \citep[using the quasar SED templates of][]{Shang2011} then our total IR luminosity is at most overestimated by a factor of 3.5 for these four IRAGN.  However, we consider this highly unlikely because the MIPS-to-IRAC flux ratios for these four IRAGN (F$_{24\micron}$/F$_{8.0\micron} = 4.2$ median) is highly discrepant from the optically luminous quasars of \citet{Shang2011} (F$_{24\micron}$/F$_{8.0\micron} = 1.4$ median) and so this factor of 3.5 is a very conservative limit for this extreme case.  We believe our estimates of the total IR luminosity to be valid due to the majority of our sample being LIRGs and having modest redshifts, based on the work of \citet{Rujopakarn2012}.

\begin{figure}
\epsscale{1.2}
\plotone{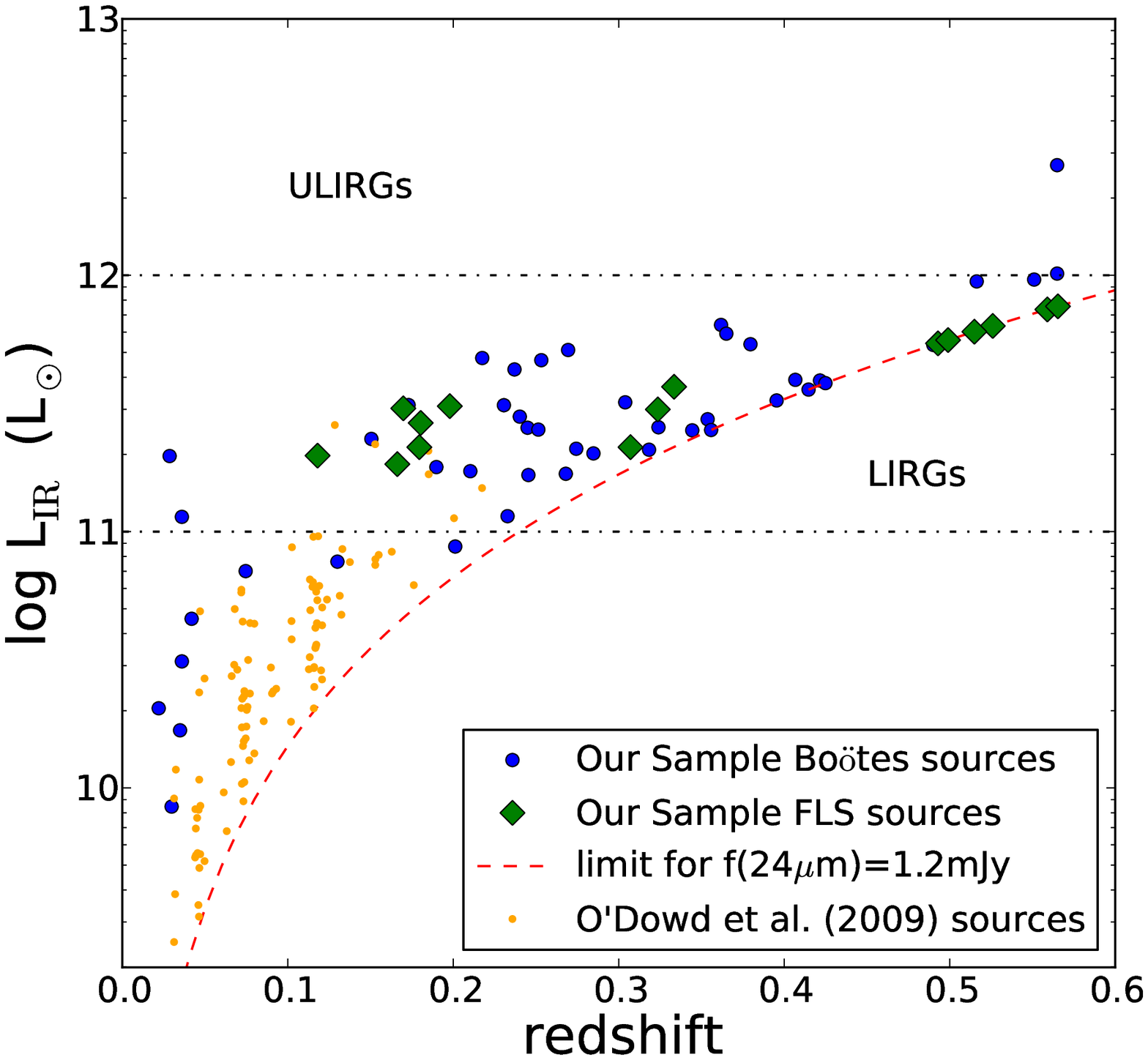}
\caption{Redshift versus the total IR luminosity from 8-1000$\micron$ for our IRS sample (blue circles, Bo\"{o}tes sources; green diamonds, FLS sources) derived from the MIPS 24~\micron\ data (and 70 and 160~\micron, if available) derived using the \citet{Rieke2009} IR SEDs.  The galaxies in our IRS sample span the range of IR luminosity of ``Luminous IR galaxies'' (LIRGs), \lir\ = $10^{11} - 10^{12}$\lsol.  The figure also shows the \citet{ODowd2009} SSGSS sample (orange circles), which are lower redshift and IR luminosity.   The dashed curve shows the limiting IR luminosity as a function of redshift for a fixed 24~\micron\ flux density of 1.2 mJy using the IR SEDs from \citet{Rieke2009}.}
\label{total IR}
\end{figure}

Figure \ref{total IR} shows the calculated total IR luminosities from fits to the MIPS data using the \citet{Rieke2009} IR SEDs.  Table \ref{table galaxy stats} lists the MIPS flux densities for 24$\micron$, 70$\micron$, and 160$\micron$, along with the total IR luminosities.  The figure also shows the relation for the IRS sample of the \spitzer\ SDSS Galaxy Spectroscopic Survey (SSGSS) from \citet{ODowd2009}, which illustrates the distinction in IR luminosity and redshift range between their sample and our sample here.  

\subsection{Contribution of PAH Emission to L$_{\mathrm{IR}}$}
\label{PAH to LIR}

\begin{figure}
\epsscale{1.2}
\plotone{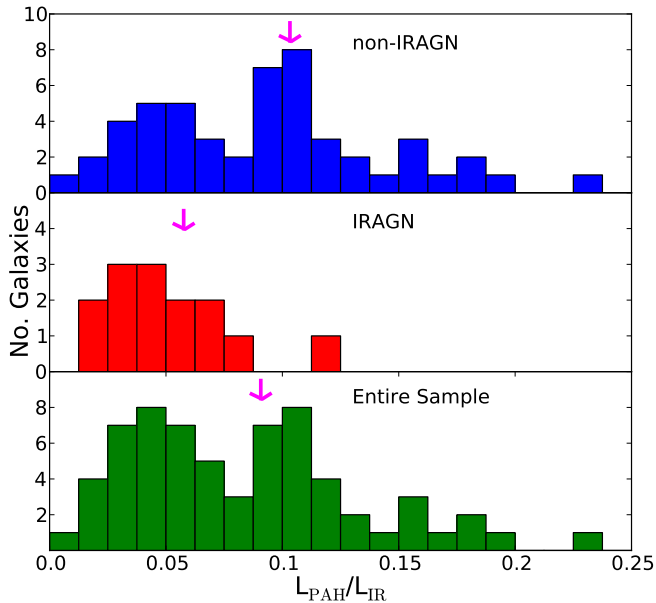}
\caption{The distribution of \lpah/\lir\ for our
sample, where \lpah\ is the total luminosity of the 6.2,
7.7, 8.6, 11.3, 12.7, and 17.0\micron\ PAH emission features.   The bottom
panel shows the distribution for our IRS sample.  The middle panel
shows the distribution for our subsample of IRAGN, and the top panel
shows the distribution for our subsample of non-IRAGN.  The median \lpah/\lir\ is 0.05 for the IRAGN,  and is about half that
for the non-IRAGN which have a median of \lpah/\lir\ $=$ 0.09.  The arrows represent the \lpah/\lir\ ratios from the composite spectra in figure \ref{composite spec} (\lpah/\lir\ = 0.09, 0.06, and 0.10 for the IRS sample, IRAGN, and non-IRAGN, respectively).}
\label{PAH contribution}
\end{figure}

In figure \ref{PAH contribution}, we show the distribution of
\lpah/\lir\ for the entire sample (bottom
panel), the subsamples of IRAGN (middle panel) and non-IRAGN (top
panel).    Here \lpah\ is the sum of all the
luminosities of the PAH features at 6.2, 7.7, 8.6, 11.3, 12.7, and
17.0~\micron, and \lir\ is the total IR luminosity as described in section \ref{LIR} using the \citet{Rieke2009} templates.  The median ratio of \lpah/\lir\ is 0.08 with an interquartile range (which contains the inner 50\% of galaxies) of 0.05 to 0.11.   We find that \lpah/\lir\ is higher in the non-IRAGN sample, with a median of 0.09, compared to \lpah/\lir\ $ =$ 0.05 for the IRAGN.  A K-S test applied to the distributions gives a D-statistic = 0.52, which indicates a 99.7$\%$ likelihood that the IRAGN and non-IRAGN have different parent distributions.  If we exclude the four IRAGN only detected at 24\micron\ and not in the far-IR, the K-S test gives a D-statistic = 0.49 and a likelihood of 97.6$\%$.

The 7.7\micron\ feature for either IRAGN or non-IRAGN contributes significantly to the entire amount of PAH emission.  The median L$_{\mathrm{7.7}}$/\lpah\ is 0.45 with an interquartile range (which contains the inner 50\% of galaxies) of 0.40 to 0.48 for our sample; and the median IRAGN L$_{\mathrm{7.7}}$/\lpah\ is slightly higher at 0.49.  These results show that as much as 10$\%$ of the total IR luminosity can come from the 7.7\micron\ feature alone.

The IRAGN in our sample show lower \lpah/\lir\ compared to the non-IRAGN, where 13/14 IRAGN have \lpah/\lir\ $<$ 0.09.  This is similar to the findings of \citet{Smith2007}, who suggested there may be a natural limit to the
absolute PAH strength in sources with AGNs, either because of partial destruction of the grains or an increase to the total infrared luminosity from other continuum sources.  However, there are two key differences between our sample and that of \citet{Smith2007}.  First, the objects in our sample have (in most cases) significantly higher IR luminosities.  Second, the IRS apertures encompass the integrated light from the galaxies in our sample, whereas the IRS resolves those in the \citet{Smith2007} sample to include light separately from the disk and nucleus.  This will tend to dilute the effects of the AGN on the integrated PAH ratios.  We discuss this further in section \ref{AGN effects}. 

\subsection{Detection Frequency of Emission Features}
\label{emission freq}
\begin{figure}
\epsscale{1.2}
\plotone{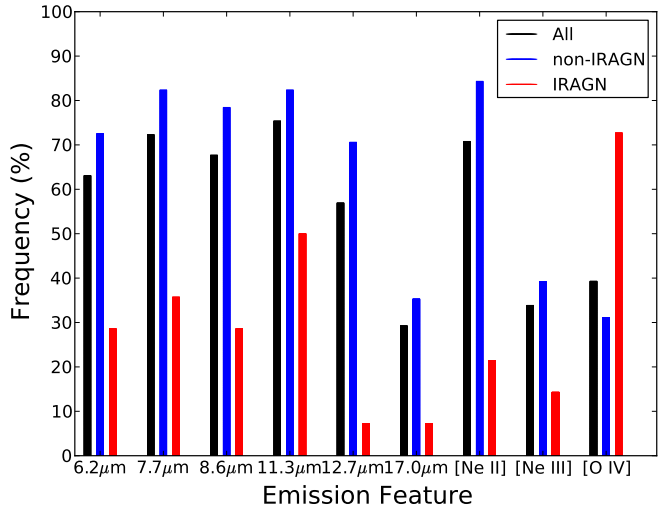}
\caption{Detection frequencies for common mid-IR PAH features and atomic emission lines in the IRS sample.    Each histogram gives the frequency that a given emission feature is detected at $\geq 3\sigma$ significance in the IRS sample (black bars), or the IRAGN (red bars) or the non-IRAGN (blue bars) subsamples.  On the abscissa, atomic features are labled, and PAH emission features are denoted by their central wavelengths.}
\label{emission frequency}
\end{figure}

Figure \ref{emission frequency} shows the fraction of galaxies in our sample whose IRS spectra show some of the most common emission features detected at $\ge 3\sigma$ significance.   Emission frequencies are given in table \ref{efreq} for the full sample, non-IRAGN, and IRAGN subsamples.  As an example, the 7.7$\micron$ PAH emission feature is found with $\ge 3\sigma$ significance in only 5 out of the 14 IRAGN in our the sample, and therefore its frequency is 36$\%$.  Although a variety of factors can influence detection of the PAH features, since our sample is selected primarily on mid-IR flux, to first order we should reach similar detection limits on average for all the sample members.

The 6.2\micron, 7.7\micron, 8.6\micron, 11.3\micron, and 12.7\micron\ PAH features are present in more than 50\% of the galaxies in our sample (and the same PAH features are present in more than 70\% of the non-IRAGN galaxies).  The most commonly detected PAH emission feature is the 11.3$\micron$ feature with a frequency of 75$\%$.  The PAH feature with the lowest detection frequency (29$\%$) is
the 17\micron\ feature.  Considering the two subsamples, the frequency of
the PAH emission features for the IRAGN subsample is less than half that
of the non-IRAGN subsample except for the 11.3$\micron$ feature, which
has 50\% for IRAGN and 82\% for the non-IRAGN.
This supports the conclusion from \citet{Diamond-Stanic2009} that the
11.3$\micron$ feature is relatively unaffected by AGN.

\begin{deluxetable}{lccr}
\tablecaption{Emission Feature Detection Frequencies\label{efreq}}
\tablecolumns{4}
\tabletypesize{\footnotesize}
\tablewidth{0pc}
\tablehead{
\colhead{Feature/Line} &
\colhead{All} &
\colhead{non-IRAGN} &
\colhead{IRAGN} \\
\colhead{} &
\colhead{galaxies} &
\colhead{galaxies} &
\colhead{galaxies}
}
\startdata
6.2\micron\ & 41/65  63\% & 37/51  73\% &  4/14  29\% \\ 
7.7\micron\ & 47/65  72\% & 42/51  82\% &  5/14  36\% \\ 
8.6\micron\ & 44/65  68\% & 40/51  78\% &  4/14  29\% \\ 
11.3\micron\ & 49/65  75\% & 42/51  82\% &  7/14  50\% \\ 
12.7\micron\ & 37/65  57\% & 36/51  71\% &  1/14 \,  7\% \\ 
17.0\micron\ & 19/65  29\% & 18/51  35\% &  1/14 \,  7\% \\ 
$[$\ion{Ne}{2}$]$ & 46/65  71\% & 43/51  84\% &  3/14  21\% \\ 
$[$\ion{Ne}{3}$]$ & 22/65  34\% & 20/51  39\% &  2/14  14\% \\ 
$[$\ion{O}{4}$]^{\dag}$ & 22/56  39\% & 14/45  31\% & 8/11  73\% \\
\enddata
\tablecomments{We present our detection frequencies for the most common PAH emission features and atomic emission lines in the mid-IR for the IRS sample.  Each column gives the frequency that a given emission feature is detected at $\geq 3\sigma$ significance in the full sample, or the IRAGN, or the non-IRAGN subsamples (given as number of galaxies with detections in the sample or as a percentage of galaxies with detections compared to their respective sample, see section \ref{emission freq}).  In the far left column, atomic features are labeled, and PAH features are denoted by their central wavelengths.  $^{\dag}$9 sources in our sample have redshifts ($z \gsim 0.5$) for which [\ion{O}{4}] is shifted out of the IRS LL channel and are not included in the sample for this line.}
\end{deluxetable}

Figure \ref{emission frequency} shows the frequency that the [\ion{Ne}{2}] $\lambda 12.8\micron$,
[\ion{Ne}{3}] $\lambda 15.6\micron$, and [\ion{O}{4}] $\lambda 25.9\micron$ emission lines are detected with $\geq
3\sigma$ significance in our IRS sample.  The emission line of [Ne II] is a tracer of the SFR \citep{Ho2007}, while the ratio of [\ion{Ne}{3}]/[\ion{Ne}{2}] and the [\ion{O}{4}] emission line trace harder ionization
parameters that correlate with AGN intrinsic luminosity
\citep[e.g.][]{Diamond-Stanic2009, Melendez2008, Rigby2009}.  The [\ion{Ne}{2}] line has a
$>$80\% detection frequency for the non-IRAGN subsample, significantly
higher than the $\sim$20\% detection frequency for the IRAGN
subsample.  In contrast, the [\ion{O}{4}] line has $>$70\%
detection frequency in the IRAGN subsample compared to $\sim$30\% for the
non-IRAGN.   These detection frequencies of [\ion{Ne}{2}] and
[\ion{O}{4}] are independent of the selection of these subsamples
(which depend solely on their IRAC colors), yet give additional
evidence that the emission from these samples stems from
star-formation (in the case of the non-IRAGN) and AGN (in the case of
the IRAGN).   We discuss the inferred physics based on the
[\ion{Ne}{2}] and [\ion{O}{4}] emission lines in Sections \ref{Ne
II O IV comparison} and \ref{oiv discussion}.

\subsection{Measures of Grain Sizes and Ionization State}
\label{grain size ion state}
Models show that the relative power emitted in the different PAH
emission features depends on the distribution of molecular grain sizes contributing to that PAH feature \citep{DraineLi2001, Schutte1993, Tielens2005}.  Generically,
the models predict that PAH molecules of smaller size should emit more
power at shorter wavelengths and PAH molecules of larger size should
emit more power at longer wavelengths (i.e., the emission from the
6.2\micron\ PAH feature should be dominated by grains of smaller
size than those at 11.3\micron).  Therefore, the ratio of the strength of longer--wavelength PAH features to shorter--wavelength
PAH features constrains the size of the PAH molecules.

In addition, the models predict that the ionization state of PAH
molecules affects the relative luminosity in each PAH emission feature
\citep{DraineLi2001}.  For example, the C-C vibrational
modes are expected to be more intense in ionized PAH molecules
\citep{Tielens2005}.  The 6.2$\micron$ and 7.7$\micron$ bands result
from radiative relaxation of C-C stretching modes and the features
should change little relative to each other as the ionization fraction
changes \citep{ODowd2009}.  In contrast, the power of the PAH bands
attributed to C-H modes such as the 11.3$\micron$
and 12.7$\micron$ features is expected to decrease (moving from
neutral to ionized clouds in the interstellar medium (ISM) ).
Therefore, the ratio of the strength of  ionization--independent PAH
features, such as the 6.2 and 7.7~\micron\ feature, to
ionization--dependent PAH features, such as the 11.3 and 12.7~\micron\
features should be sensitive to the overall ionization state of the
PAH molecules.  

\begin{figure}
\epsscale{1.2}
\plotone{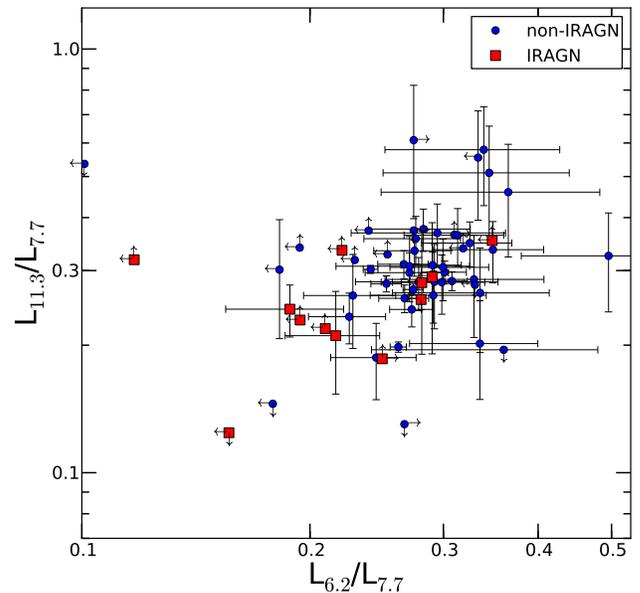}
\caption{PAH band ratios L$_{6.2}$/L$_{7.7}$ vs. L$_{11.3}$/L$_{7.7}$
for the galaxies in the IRS sample.  The L$_{6.2}$/L$_{7.7}$ ratio is a tracer of PAH grain size, and the
L$_{11.3}$/L$_{7.7}$ ratio is a tracer of PAH grain ionization
\citep{ODowd2009}.  The red squares denote IRAGN and the blue circles
denote non-IRAGN galaxies.  The IRS sample falls mostly within the locus
defined by 0.2--0.4 in both axes.  This shows that the galaxies have a
mixture of grain sizes and different ionization fractions of PAHs.  We
measure a weak correlation between L$_{6.2}$/L$_{7.7}$ and
L$_{11.3}$/L$_{7.7}$ (Spearman's $\rho =$ 0.41), which we interpret as
evidence for decreasing PAH grain size and increasing ionization within the locus.
We also observe that the IRAGN ratios are slightly lower typically compared to the non-IRAGN galaxies, in our IRS sample.}
\label{Grain size Ionization state}
\end{figure}

Therefore, the PAH molecule sizes and ionization state can be
constrained by comparing the PAH band ratios L$_{6.2}$/L$_{7.7}$ to
L$_{11.3}$/L$_{7.7}$.  The former traces PAH grain size, but is
relatively unaffected by ionization.  The latter is sensitive to ionization and less affected by changes in grain size.
Figure \ref{Grain size Ionization state} shows these line ratios for the galaxies in our sample.  The sample lies mostly in a locus between line ratios of 0.2--0.4 for both L$_{6.2}$/L$_{7.7}$ and L$_{11.3}$/L$_{7.7}$.  This is consistent with the findings from other studies \citep{ODowd2009, Diamond-Stanic2010, Wu2010}.  We measure a weak correlation between L$_{6.2}$/L$_{7.7}$ and L$_{11.3}$/L$_{7.7}$ (Spearman's $\rho =$ 0.41), which we interpret as evidence for decreasing PAH grain size with increasing ionization within the locus \citep[this is consistent with the trend seen in][]{ODowd2009}.

\subsection{Relation between Radiation Hardness and PAH Strength}
\label{radiation hardness}
\begin{figure}
\epsscale{1.2}
\plotone{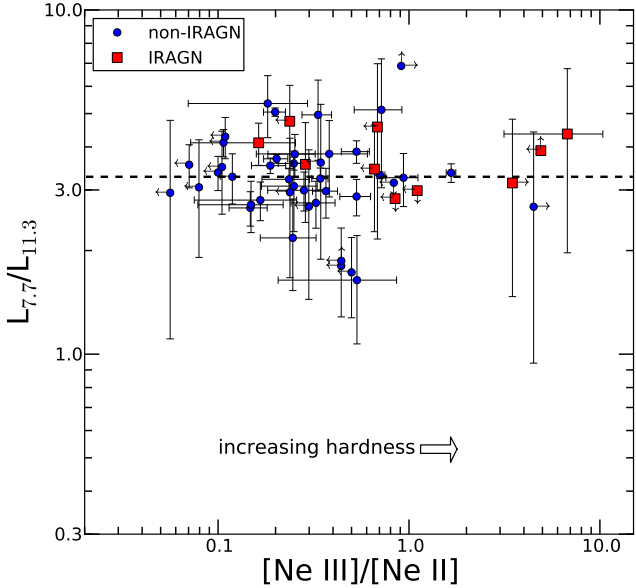}
\caption{[\ion{Ne}{3}]$_{15.6\mu m}$/[\ion{Ne}{2}]$_{12.8\mu m}$ versus
L$_{7.7}$/L$_{11.3}$.   Blue circles show the subsample of non-IRAGN,
and red squares show the subsample of IRAGN (not all IRAGN have [\ion{Ne}{3}]/[\ion{Ne}{2}] ratio detections).  The dashed line is the
median L$_{7.7}$/L$_{11.3}$ value for the non-IRAGN subsample
only. There is no trend between [\ion{Ne}{3}]/[\ion{Ne}{2}] and
L$_{7.7}$/L$_{11.3}$ for either the non-IRAGN or IRAGN. This is
similar to the conclusions from other studies \citep{ODowd2009,
Wu2010}, but contrasts with the findings of \citet{Smith2007},
although we expect this difference, in part, is due to issues with the full integrated
light \citep[nucleus and outer regions from the galaxies in our IRS sample fall
within the IRS aperture in constrast to the sample of][]{Smith2007}.}
\label{Ne ratio}
\end{figure}

The emission-line ratio of the [\ion{Ne}{3}] (15.6$\micron$ with an
ionization potential of 41eV) emission to the  [\ion{Ne}{2}] (12.8
\micron, with an ionization potential of 21.6eV) is a measure of the
hardness of the radiation field \citep{Smith2007}.  Figure \ref{Ne
ratio} shows the ratio of L$_{[\mathrm{Ne III}]}$/L$_{[\mathrm{Ne
II}]}$ as a function of L$_{7.7\micron}$/L$_{11.3\micron}$, which is
an indicator of the ratio of ionized-to-neutral PAH molecules, as
discussed in section \ref{grain size ion state}.  In general, there is no correlation between L$_{[\mathrm{Ne
III}]}$/L$_{[\mathrm{Ne II}]}$ and L$_{7.7\micron}$/L$_{11.3\micron}$.
The dashed line in figure~\ref{Ne ratio} shows the median
L$_{7.7\micron}$/L$_{11.3\micron}$ for the non-IRAGN only.  There is tentative evidence that the IRAGN have L$_{7.7\micron}$/L$_{11.3\micron}$ ratios higher than this median, but this is not statistically significant given our sample size.

We see no IRAGN galaxies with extremely low  L$_{7.7\micron}$/L$_{11.3\micron}$ ratios ($<$1), as seen in the \citet{Smith2007} sample.  The L$_{7.7\micron}$/L$_{11.3\micron}$ ratio behavior contrasts strongly with \citet{Smith2007}, but is in reasonable agreement with others \citep{Wu2010, ODowd2009}.  We suspect the difference is a result of the fact that the IRS spectra for our sample contains the integrated light from the galaxy (nucleus + galaxy),
whereas the IRS apertures containing the nuclei used by \citet{Smith2007} included only the galactic nuclear
regions.  Our sample contains few sources with high hardness ratios \citep[$\mathrm{[Ne III]/[Ne II]}$
$>$ 2, similar to the samples of][]{ODowd2009, Smith2007, Wu2010}, and most sources with high hardness ratios are upper limits only.  This suggests that higher hardness ratios cannot account for the trend seen by
\citet{Smith2007}.  This conclusion agrees with that in \citet{Gordon2008} from a study of H II regions in M101 and \citet{Brandl2006} studying 22 starburst galaxies.  \citet{Diamond-Stanic2010} showed that spectra taken from an off-axis region of an AGN host galaxy have no significant
differences from SINGS H II galaxies' PAH ratios.  As discussed in
Section \ref{emission freq}, the 11.3$\micron$ feature is relatively
unaffected by AGN, which could explain the lower ratios seen by
\citet{Smith2007}.  The other studies also have samples that would remain
relatively unchanged over their respective ranges of hardness ratios
\citep[z$_{\mathrm{med}}$ = 0.08 and normal SF galaxies and 0.008 $<$
z $<$ 4.27, respectively]{ODowd2009, Wu2010}.

\subsection{The Distribution of the 6.2$\micron$ PAH Equivalent Width}

Previous studies use the equivalent width (EW) of the 6.2$\micron$ PAH
feature to classify sources as AGN, starbursts (SBs), and composite
AGN+SB subsamples \citep[e.g.,][]{Wu2010}.  Sources where an AGN dominates the IR emission are expected to have
lower 6.2\micron\ PAH EW either because the mid-IR continuum from the
AGN is much stronger and washes out the emission, or because the
ionization field of the AGN destroys the PAH molecules, or a
combination of the two.  The EW measures the relative strength of the PAH feature to the mid-IR continuum, thereby allowing us to quantify the effects of an AGN.

\begin{figure}
\epsscale{1.2}
\plotone{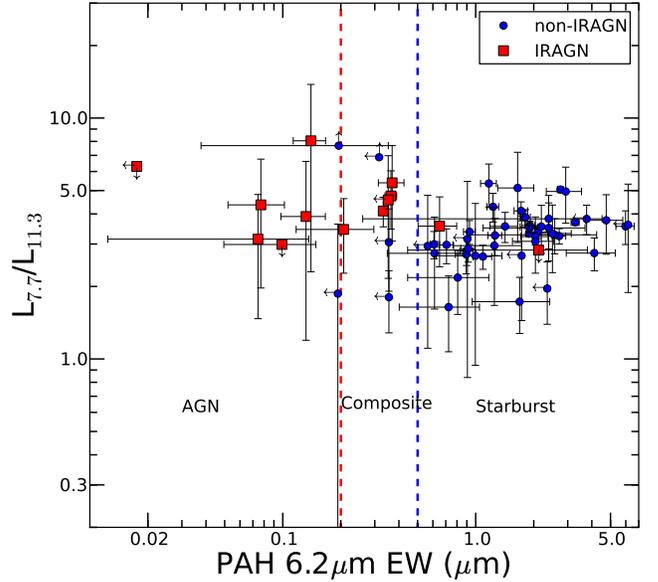}
\caption{PAH 6.2\micron\ equivalent width versus the L$_{7.7}$/L$_{11.3}$ PAH ratio.  The dashed lines show selection criteria
to separate sources into AGN, composite, and SB galaxies, as labeled
in the figure.  Galaxies with EW $\leq$ 0.2 $\micron$ are likely AGN
(red-dashed line), galaxies with EW $>$ 0.5 $\micron$ are SB galaxies
(blue-dashed line), and galaxies that fall between these lines are
likely AGN+SB composites.  We observe no correlation between
L$_{7.7}$/L$_{11.3}$ and the EW of the 6.2$\micron$ PAH feature.}
\label{EW62}
\end{figure}

Figure \ref{EW62} shows the PAH 6.2\micron\ EW versus the
7.7--to--11.3\micron\ PAH flux ratio.   The red dashed line shows a
selection for AGN with EW $\leq$ 0.2$\micron$, the blue dashed line
shows a selection for SB galaxies with EW $>$ 0.5$\micron$, and AGN+SB
composite sources populate the region between these regions
\citep{Armus2007}.  Most of the IRAGN (defined by the IRAC color-color selection) have low
6.2$\micron$ EW:  12 of the 14 IRAGN satisfy either the composite or
AGN 6.2\micron\ PAH EW criteria.   Similarly, most of the non-IRAGN
have high 6.2\micron\ EW:  46 of 51 non-IRAGN satisfy the SB 6.2\micron\ PAH
EW criterion.  Interestingly, the figure shows no correlation
between the 6.2$\micron$ EW and L$_{7.7\micron}$/L$_{11.3\micron}$ ratio.  A similar observation is shown for the L$_{7.7\micron}$/L$_{11.3\micron}$ ratio compared to the hardness of the radiation field in figure \ref{Ne ratio} (see section \ref{radiation hardness}).  Therefore, we see no indication that the presence of an AGN affects the L$_{7.7\micron}$/L$_{11.3\micron}$ ratio.  If the AGN were photoionizing the PAH molecules; and reducing the 6.2\micron\ EW, we would expect the L$_{7.7\micron}$/L$_{11.3\micron}$ ratio to decrease with decreasing 6.2\micron\ EW.  Because we do not, we conclude that the AGN decreases the PAH 6.2\micron\ EW because the AGN increases the mid-IR continuum, and not because the AGN destroys the PAH molecules on a galaxy wide scale (see section \ref{short to long}).

\subsection{The relationship between PAH luminosity, Star-formation Rate, and AGN luminosity}
\label{Ne II O IV comparison}

\begin{figure}
\epsscale{1.2}
\plotone{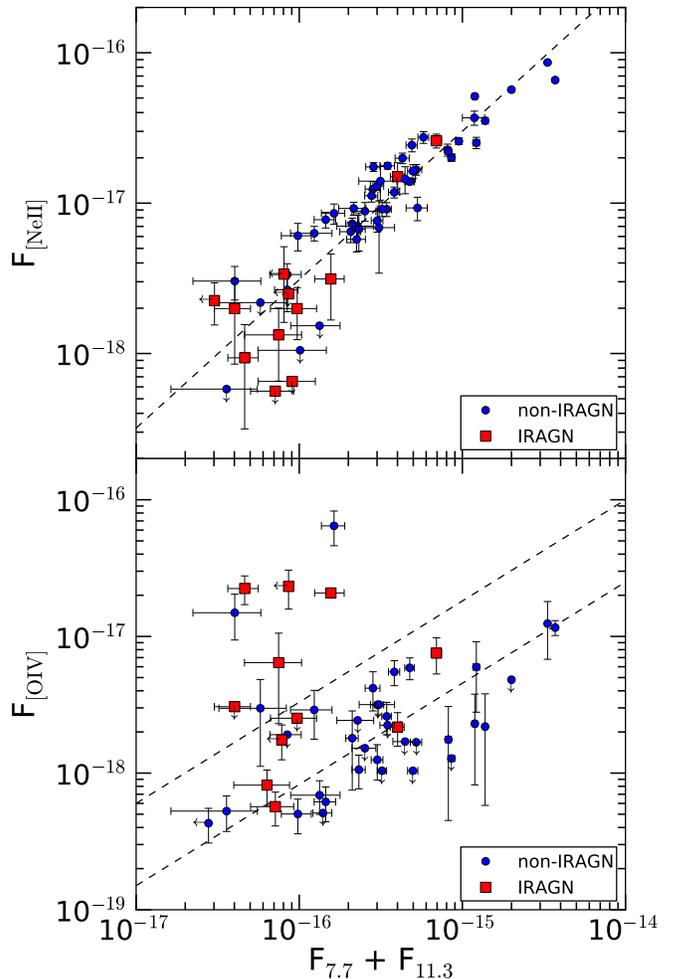}
\caption{The relationship between the PAH emission features and the
[\ion{Ne}{2}] (top panel) and [\ion{O}{4}] (bottom panel) emission
lines.  All units are in W/m$^2$.  The top panel shows a correlation
between the PAH emission and [\ion{Ne}{2}] emission line suggesting PAH
emission correlates with star formation.  The dashed line represents a linear fit (0.02$\times$F$_{7.7+11.3}$$^{0.99}$) to the sample.  The bottom panel shows that there is little, if any, correlation between
PAH emission and [\ion{O}{4}], but there is a population of objects
with low PAH emission and excess [\ion{O}{4}] emission.  The upper dashed line is four times greater than the fit to the non-IRAGN subsample (lower dashed line) used to select the [\ion{O}{4}]-excess objects.}
\label{Ne II O IV}
\end{figure}

As discussed above, the [\ion{Ne}{2}] and the [\ion{O}{4}] emission lines are useful probes of the SFR and AGN luminosity, respectively.  In figure \ref{Ne II O IV}, we plot the flux measured in the [\ion{Ne}{2}] and [\ion{O}{4}] lines against the summed flux from the 7.7\micron\ and 11.3\micron\ PAH features, which we are able to then compare directly against results from \citet{Diamond-Stanic2010}.  The [\ion{Ne}{2}] emission correlates strongly with PAH emission, with a Spearman correlation coefficient of $\rho = $ 0.94 and a linear fit of F([\ion{Ne}{2}]) = 0.02 $\times$F$_{7.7+11.3}$$^{0.99}$.   The scatter in the [\ion{Ne}{2}]--PAH relation increases for fainter fluxes, which may be related to the presence of IRAGN, which are more frequent in the sample at faint line fluxes.

In contrast, the [\ion{O}{4}] line shows almost no
correlation with PAH emission, with a Spearman correlation coefficient
of $\rho = $ 0.19.  Because [\ion{Ne}{2}] correlates so well with the
PAH emission, the relation in figure~\ref{Ne II O IV} suggests that
the PAH emission correlates with the SFR.  This is consistent with the
findings of \citet{Diamond-Stanic2010} and \citet{Genzel1998}.  We also do not see any clear
difference between the IRAGN and non-IRAGN in either the [\ion{Ne}{2}]
or [\ion{O}{4}] versus PAH emission plot in figure~\ref{Ne II O IV},
indicating further that the PAH features are likely tracing star
formation in both subsamples, and that the presence of AGN do not
increase (or decrease) PAH emission integrated over galaxies
\citep{Diamond-Stanic2010}.

The bottom panel in figure~\ref{Ne II O IV} shows a swath of eight
galaxies with low PAH emission and high [\ion{O}{4}] emission.   These
galaxies lie above the upper dashed line, which is 4 times greater than
the fit to the non-IRAGN (lower dashed line).  Approximately one-half (5/11\footnote{Three do not have [\ion{O}{4}] detections due to the redshift of the source.}) of the IRAGN have [\ion{O}{4}] emission above
this line.  In addition, three  of the galaxies above this line are
non-IRAGN.  Removing these galaxies from the sample shows a correlation may be present (Spearman's $\rho = $ 0.60) for galaxies without strong [\ion{O}{4}] emission.  As stated in Section \ref{AGN selection}, the
\citet{Stern2005} IRAC color-color method used to identify IRAGN
selects more broad-line AGN, but only about 40$\%$ of narrow-lined
AGN.  Because excess [\ion{O}{4}] indicates the presence of ionization
from an AGN, these three non-IRAGN galaxies likely harbor
heavily-obscured AGN that are missed by the IRAGN selection.

\vspace{0.75cm}
\section{Discussion}
\label{discussion}

\subsection{The Color-Magnitude Diagram for IR Luminous Galaxies}
\label{LIR CMD}

\begin{figure}
\epsscale{2.4}
\plottwo{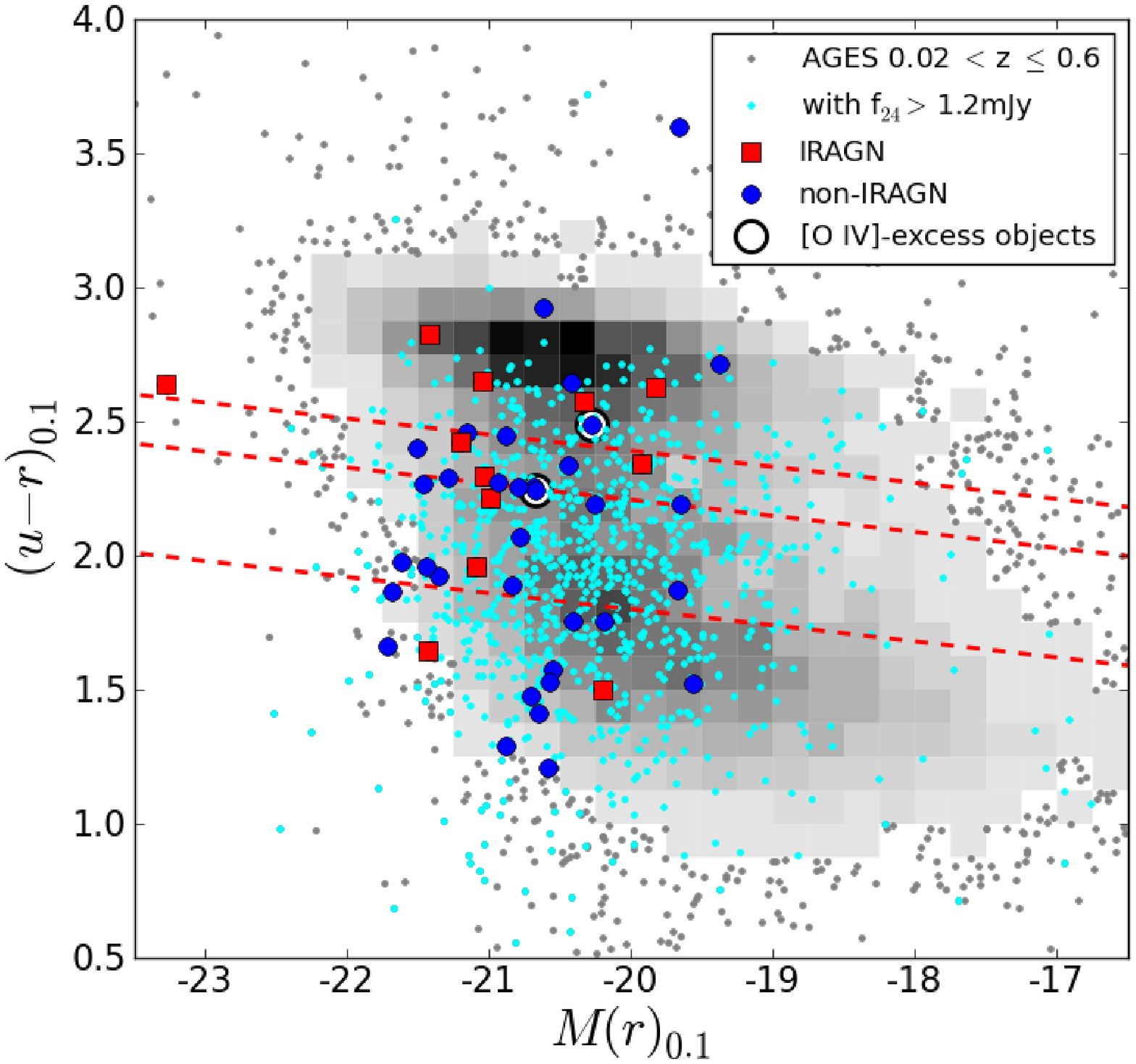}{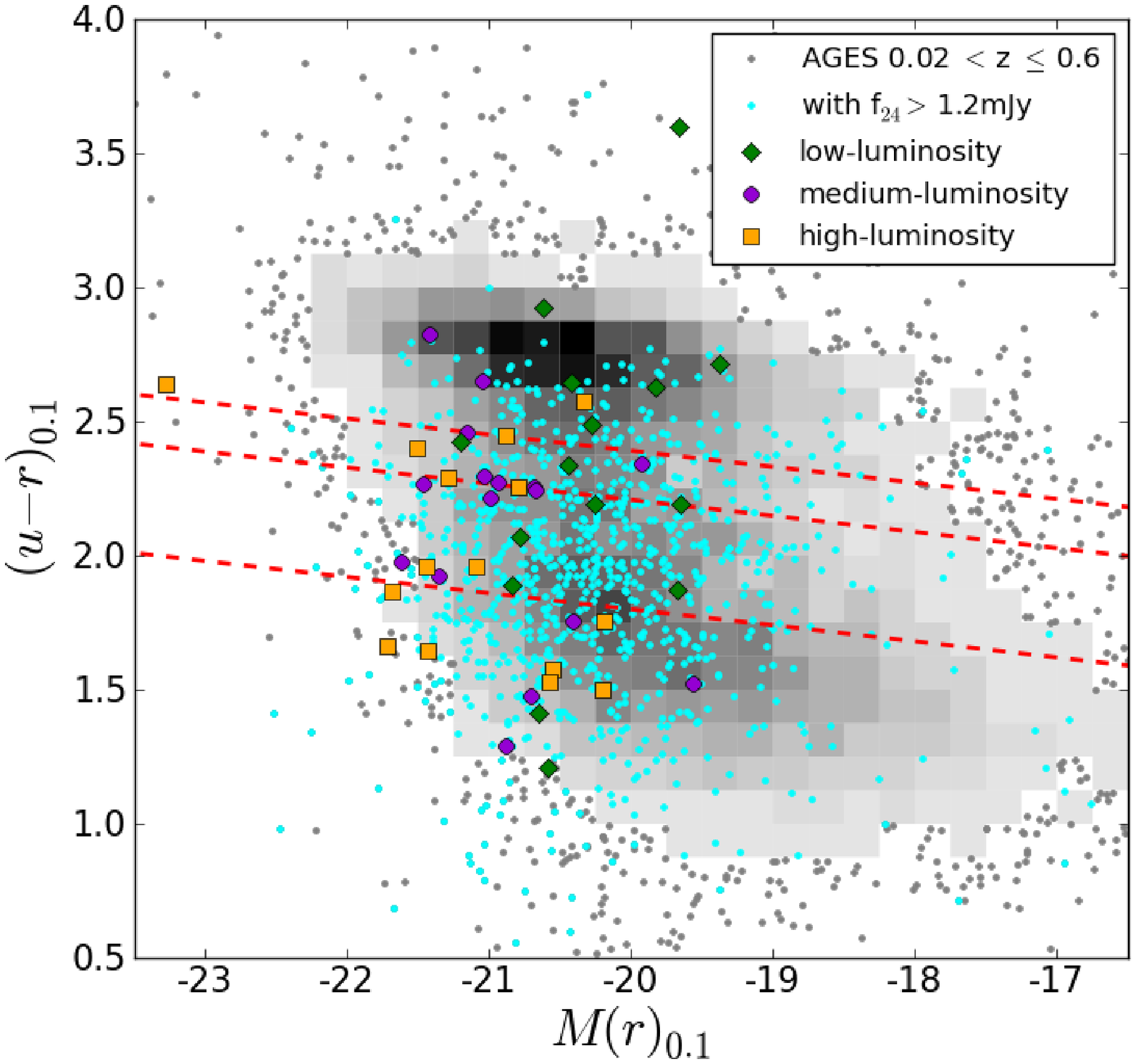}
\caption{Optical $(u-r)_{0.1}$ color-magnitude diagram.  Both plots are similar to figure \ref{CMD} except for symbol changes to the Bo\"{o}tes IRS sample.  (\textit{Top Panel}) The sample symbols denote IRAGN (red squares), non-IRAGN (blue circles), and non-IRAGN galaxies with excess [\ion{O}{4}] (defined in section \ref{Ne II O IV comparison}, black open circles).  (\textit{Bottom Panel})  The sample symbols denote galaxy IR luminosity as low-luminosity galaxies (L$_{\mathrm{IR}} = 10^{10} - 2.4 \times 10^{11}$
L$_{\odot}$, green diamonds), medium-luminosity galaxies ($\lir = 2.4 \times 10^{11} - 4.5 \times 10^{11}$
L$_{\odot}$, purple circles), and high-luminosity galaxies (L$_{\mathrm{IR}} > 4.5 \times 10^{11}$L$_{\odot}$, orange squares).}
\label{IR CMD}
\end{figure}

Because our IRS sample spans a nearly uniform distribution in optical color, we are able to study differences in the IR emission properties of galaxies with different positions on the optical color-magnitude relation.  For the full sample (all galaxies in AGES with $f(24\micron) > 1.2$~mJy and 0.02 $< z <$ 0.6, including both the IRAGN and non-IRAGN), we find the median values are $M(r)_{0.1}$ = -20.36 mag and $(u-r)_{0.1}$ = 1.97 mag.  These differ from the median values for our IRS sample, which are $M(r)_{0.1}$ = -20.74 mag and $(u-r)_{0.1}$ = 2.23 mag).  We attribute these differences to the fact that our IRS sample has a uniform selection in $(u-r)_{0.1}$ optical color whereas the full sample is weighted toward bluer $(u-r)_{0.1}$ optical colors.  In figure \ref{IR CMD}, we replot the location of the IR-luminous galaxies on the optical $M(r)_{0.1}$ versus $(u-r)_{0.1}$ color-magnitude diagram, and we denote galaxies by their IR luminosity as derived from their MIPS data (see section \ref{LIR}) in coarse bins of luminosity:  low-luminosity galaxies ($ \lir = 10^{10} - 2.4 \times 10^{11}$~\lsol), medium-luminosity ($\lir = 2.4 \times 10^{11} - 4.5 \times 10^{11}$~\lsol) and high-luminosity ($\lir > 4.5 \times 10^{11}$~\lsol), evenly splitting the medium and high luminosity bins ($\sim$100 galaxies for each bin) for the full sample to improve the statistics for comparison\footnote{If we use the traditional ULIRG identification ($\lir~> 10^{12}\lsol$) for the full sample, we have only 9 galaxies for comparison.}.  Using the full sample of galaxies with $f(24\micron) > 1.2$~mJy, we find no difference in $(u-r)_{0.1}$ color as a function of IR luminosity.  Each of the three IR luminosity bins have nearly equal median colors $(u-r)_{0.1}$ = 2.0 mag with similar interquartile ranges.  The only difference we measure is that the $(u-r)_{0.1}$ color distribution of the highest IR luminosities is slightly broader than for the lower luminosity galaxies.

In comparison, our IRS sample shows the low, medium and high-luminosity galaxies are also mostly distributed evenly in the color-magnitude relation.  We observe slight evidence that the $(u-r)_{0.1}$ color \textit{decreases} for \textit{increasing} IR luminosity.  We find the median $(u-r)_{0.1}$ colors are 2.34 mag, 2.25 mag, and 1.96 mag for the low, medium, and high-luminosity galaxies, respectively.  However, we consider this trend not to be significant given the lack of any correlation in the full sample.

We compared the $(u-r)_{0.1}$ color distributions for galaxies classified as IRAGN and non-IRAGN from the full sample of galaxies with $f(24\micron) > 1.2$~mJy from AGES (cyan points in figure \ref{IR CMD}).  The median values, $M(r)_{0.1}\sim$ -20.34 mag and $(u-r)_{0.1}\sim$ 2.00 mag for the IRAGN and for the non-IRAGN $M(r)_{0.1}\sim$ -20.36 mag and $(u-r)_{0.1}\sim$ 1.96 mag are similar.  A K-S test applied to the distributions gives a D-statistic = 0.06 and a likelihood of 74.8$\%$, which we interpret as evidence that there is no difference between the $(u-r)_{0.1}$ colors for IR luminous galaxies selected as IRAGN and non-IRAGN.

Considering only our IRS sample, the IRAGN span the range from the blue cloud to the red sequence, where 67$\%$ (8 of 12) are located in the ``green valley'' or the red sequence ($(u-r)_{0.1} + 3/50 \times (M[r]_{0.1} + 20) > 2.18$).  The distribution of $(u-r)_{0.1}$ color for the non-IRAGN galaxies appears shifted toward bluer colors compared to the IRAGN.  We find that the median $(u-r)_{0.1}$ color for the non-IRAGN is 2.13 mag compared to the median color for the IRAGN, 2.38 mag, shown in figure \ref{IR CMD}.  However, because we see no difference between IRAGN and non-IRAGN in the full sample of IR luminous galaxies, we do not consider these differences significant.

Other studies of the optical colors for AGN have found that they populate a similar region of the color-magnitude relation compared to our IRAGN samples here \citep{Nandra2007, Weiner2007}.  While our IRAGN span a similar $(u-r)_{0.1}$ color range as X-ray selected AGN \citep[e.g.][]{Nandra2007}, so do the non-IRAGN.  So it is unclear if the presence of an AGN is driving redder $(u-r)_{0.1}$ colors in our IRS sample of IR luminous galaxies.  Therefore, in galaxies selected to be IR luminous (with $f(24\micron) > 1.2$~mJy), there is no measurable difference between the optical $(u-r)_{0.1}$ colors for galaxies selected as IRAGN and non-IRAGN.

However, there is a difference in our interpretation of the $(u-r)_{0.1}$ colors for the IRAGN and non-IRAGN.  Most of the IRAGN in our IRS sample have low [\ion{Ne}{2}] and PAH luminosities, which may imply they have significantly lower SFRs compared to the non-IRAGN (figures \ref{Ne II O IV} and \ref{L Ne II O IV}).  Therefore, the fact that the IRAGN in our samples lie in the ``green valley'' or red sequence may imply they have declining SFRs or a recent cessation of star-formation, similar to conclusions reached by \citet{Nandra2007}.  This also implies that star-formation is not a necessary component for an AGN to persist.

\subsection{AGN effects on PAH Emission and AGN contribution to \lir}
\label{AGN effects}

Previous studies have used anticorrelations in the [\ion{Ne}{3}]/[\ion{Ne}{2}] versus L$_{7.7\micron}$/L$_{11.3\micron}$ ratios (see figure \ref{Ne ratio}) to conclude that galaxies with an increasing hardness of radiation field from an AGN have relatively less emission from shorter wavelength PAH features compared to longer wavelength PAH features \citep[e.g.][]{Smith2007, ODowd2009, Wu2010}.  One interpretation is that the smaller PAH molecules are preferentially destroyed in the presence of the AGN \citep{Smith2007, ODowd2009}.   It is certainly the case that AGN do not \textit{increase} PAH emission, as shown in figure~\ref{Ne II O IV} \cite[see also][]{Diamond-Stanic2010}.

\begin{figure}
\epsscale{1.2}
\plotone{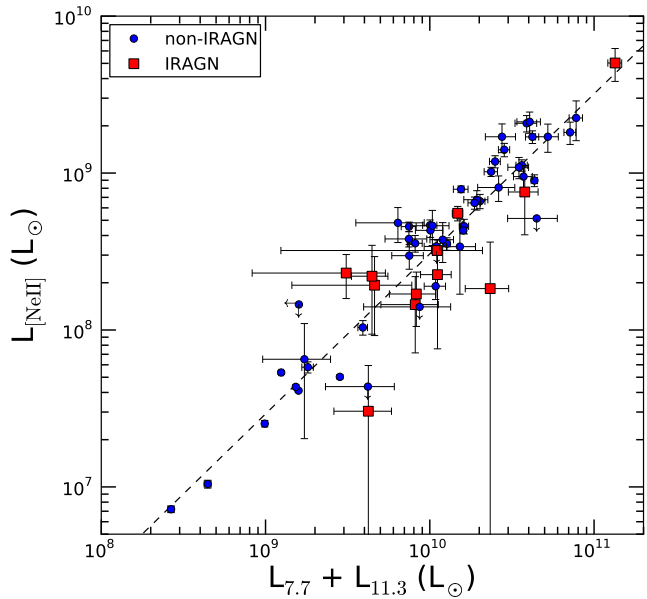}
\caption{The relationship between the luminosity of the PAH emission features and the [\ion{Ne}{2}] emission line.  The [\ion{Ne}{2}] luminosity correlates strongly with the PAH luminosity, with a Spearman correlation coefficient of $\rho$ = 0.90 and a linear fit (dashed line) of Log(L[\ion{Ne}{2}]) = 1.02 $\pm$ 0.03 $\times$ Log(L$_{7.7+11.3}$) - 1.71 $\pm$ 0.25.}
\vspace{0.25cm}
\label{L Ne II O IV}
\end{figure}

In our sample, the luminosity for all PAH features is uniformly weaker in the IRAGN galaxies
compared to the non-IRAGN and uniformly less frequent (see figures~\ref{composite spec} and 8).  Figure \ref{composite spec} shows that the mid-IR SEDs of the IRAGN in our sample show a weakly rising continuum with increasing wavelength (with possible silicate absorption at 9.7$\micron$).  Figure \ref{Ne II O IV} shows that most of the IRAGN galaxies in our sample have low [\ion{Ne}{2}] emission and PAH emission.  Therefore, the IRAGN appear to have lower implied SFRs.  
Taken together, the data suggest that the reason IRAGN in our sample have lower PAH emission is a combination of higher mid-IR continuum from processes associated with the AGN combined with intrinsically lower SFRs.  It is unclear if the AGN actually destroy PAH molecules but given the low [\ion{Ne}{2}] fluxes, it seems more likely the lack of PAH emission is because the hosts of IRAGN have lower SFRs.  It is possible that some other feedback mechanism from the AGN affects the SFR.  We also have to be cautious about the selection effect for IRAGN that the AGN will boost the flux at 24\micron\ and will select on a lower star-formation threshold than a purely star-forming sample.  Thus, this may result in weaker observed PAH emission for the IRAGN.

\begin{figure}
\epsscale{1.2}
\plotone{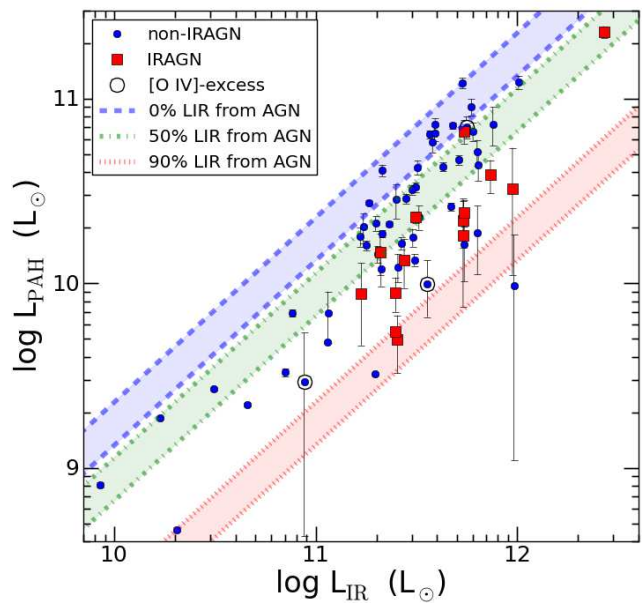}
\caption{The total IR luminosity versus the PAH luminosity as an indicator for the contribution of AGN luminosity to the \lir.  The shaded regions represent percent amounts, marked in the plot, for the contribution of AGN luminosity to the total IR luminosity for the IRAGN in our sample under the assumption that the PAH luminosity traces the total SFR (see section \ref{AGN effects} for explanation).}
\label{sf cal}
\end{figure}

Figure \ref{L Ne II O IV} shows that the PAH luminosity correlates linearly with the [\ion{Ne}{2}] luminosity (linear fit of Log(L[\ion{Ne}{2}]) = 1.02 $\pm$ 0.03 $\times$ Log(L$_{7.7+11.3}$) - 1.71 $\pm$ 0.25).  Because the correlation is linear, we conclude that both are good tracers of the total SFR \citep{Ho2007}.  We can then use the PAH luminosity and total IR luminosity to estimate the contribution of star-formation and AGN luminosity to the total IR luminosity for the IRAGN in our sample.  In figure \ref{sf cal}, we show our estimated contribution of AGN luminosity to the total IR luminosity for the IRAGN.  We make the implicit assumption that the non-IRAGN galaxies with the highest \lpah/\lir\ ratios have no contribution from AGN luminosity to their \lir.  Under this assumption the contribution of AGN to the IR luminosity is a lower limit, and could be higher by a \textit{systematic} amount.  However, the relative contribution of an AGN to the IR luminosity is robust under this assumption.  Using this premise, we define our 0$\%$ AGN luminosity contribution for the IRAGN by fitting the range of the \lpah/\lir\ ratios for the upper quartile of non-IRAGN galaxies with \lir $\geq 10^{11}$L$_\odot$ using a unity relation between the two luminosities.  The shaded region represents the range of scatter in our sample as to which galaxies are completely star-forming.  Using this definition, we attribute 50-90\% of the total IR luminosity in the IRAGN to an AGN rather than star-formation \citep[a similar result to][from a study of QSOs]{Schweitzer2006}, except for one object where the AGN contributes $< 50\%$ of the total IR luminosity (and has one of the highest [\ion{Ne}{2}] luminosities of the IRAGN).  While star-formation is a significant contribution to the total IR luminosity for the IRAGN, the AGN is the dominant source of IR luminosity in these galaxies.

\subsection{Emission ratios of Short-to-Long wavelength PAHs in IRAGN and non-IRAGN}
\label{short to long}
\citet{ODowd2009} showed that the AGN in their sample have significantly lower L$_{7.7\micron}$/L$_{11.3\micron}$ values than the SF galaxies \citep[similar to conclusions from ][]{Wu2010}.    Both O'Dowd et al.\ and Wu et al.\ found no significant difference in the L$_{6.2\micron}$/L$_{7.7\micron}$ ratio going from galaxies dominated by star-formation to those dominated by an AGN.  Taken together, these observations suggest that galaxies with AGN show a reduction in the emission of shorter wavelength PAHs relative to longer wavelength PAHs.  In contrast, we see no difference in either the  L$_{7.7\micron}$/L$_{11.3\micron}$ or L$_{6.2\micron}$/L$_{7.7\micron}$ ratio for the IRAGN and non-IRAGN in our sample.  Here we discuss possible reasons for the differences seen between the samples.  

As noted above, previous studies argued that the smaller dust grains
responsible for the shorter wavelength PAH molecules are
preferentially destroyed in the presence of an AGN \citep{Smith2007,
ODowd2009, Wu2010}.  \citet{Diamond-Stanic2010} investigated possible
destruction of PAH grains by AGN-driven shocks as X-ray heating was
shown not to be as important as shock excitation \citep{Roussel2007}.
While the effects of AGN-driven shocks on the observed PAH emission are
complex and uncertain, one possible effect is that shocks leave uneven
structures in the surviving dust grains that do not contribute to the
PAH emission.  However, \citet{Diamond-Stanic2010} noted that the
molecules responsible for the 11.3$\micron$ feature are not strongly
suppressed by shocks, and this can explain the observed lower
L$_{7.7\micron}$/L$_{11.3\micron}$ ratios.

\citet{Diamond-Stanic2010} also found that AGN-dominated sources
(the nuclei of local Seyfert galaxies) with low L$_{7.7\micron}$/L$_{11.3\micron}$
ratios have strong H$_{\mathrm{2}}$ emission (H$_{\mathrm{2}}$ S(3)
rotational line at 9.67\micron) indicating the presence of
shocks\footnote{refer to \citet{Diamond-Stanic2010} for further
description of other studies with similar results}.  This is
consistent with observations of \citet{Roussel2007} and
\citet{Smith2007} who observe stronger H$_{\mathrm{2}}$ emission and
lower L$_{7.7\micron}$/L$_{11.3\micron}$ ratios in local galaxies with
AGN from SINGS.  We do not have many sources with detected H$_{\mathrm{2}}$ emission in the 9.67\micron\ line (10/65 galaxies have H$_{\mathrm{2}}$ flux detections at $\geq$ 3$\sigma$).  This suggests that, if shocks are present, they may not be strong enough to modify the short-to-long PAH ratios observed or to produce strong emission in this H$_{\mathrm{2}}$ line.  

As our sample has the highest median redshift (z$_{\mathrm{med}}$ = 0.28) of other recent studies \citep[z$_{\mathrm{med}}$ = 0.08 and z$_{\mathrm{med}}$ = 0.144, respectively]{ODowd2009, Wu2010} we view more integrated light from a galaxy and this may result in a significant difference not associated with the effects of an AGN.  The discussion above only takes into account the differences seen between active galaxies and SF galaxies in short-to-long PAH ratios.  As \citet{ODowd2009} discuss, the age and evolution of metallicity in a galaxy can also have an effect on the observed short-to-long PAH ratios.

\vspace{0.5cm}
\subsection{Galaxies with Excess [O IV] $\lambda$25.9\micron\ Emission}
\label{oiv discussion}

Eight galaxies in our IRS sample show unusually high [\ion{O}{4}] emission relative to their PAH luminosities (figure \ref{Ne II O IV}), including three galaxies classified as non-IRAGNs (IDs  6, 50, and 55).  The excess [\ion{O}{4}] indicates the presence of a hard ionizing field, mostly likely from an AGN that is otherwise undetected in the IR.  Here we discuss these ``[\ion{O}{4}]-excess'' objects in detail.

\begin{figure}
\epsscale{1.2}
\plotone{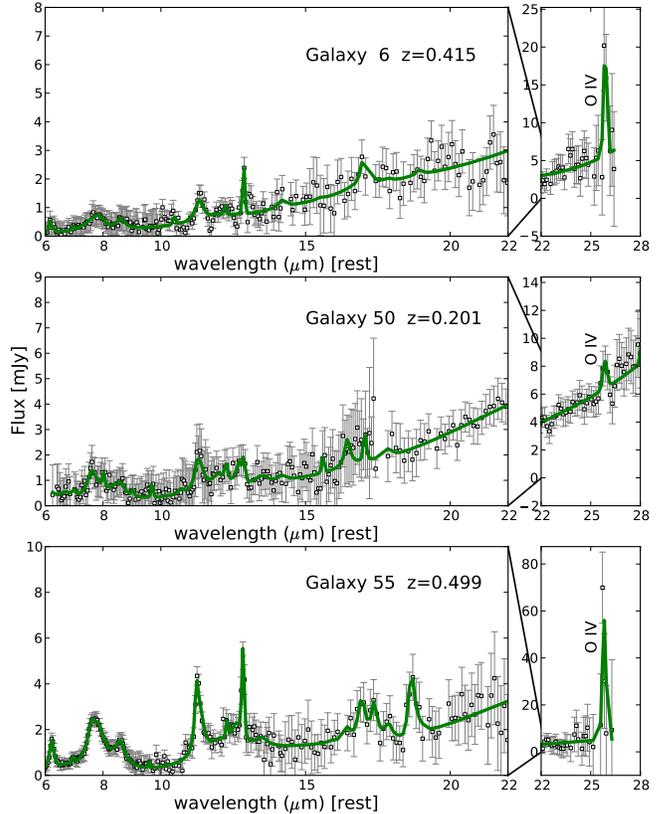}
\caption{The IRS mid-IR spectra of three non-IRAGN galaxies in our IRS sample with excess [\ion{O}{4}] emission and no other indications of an AGN.  The continuous spectrum of each galaxy is shown, but split to better show the spectrum and excess [\ion{O}{4}] emission.  Note range of y-axis differs from left to right panel for each spectrum.  The open circles are the measured flux densities from the \spitzer\ IRS data observations with error bars representing the
uncertainties in those measurements and the green line represents the
best fit from spectral decomposition of PAHFIT.  All three galaxies
show clear emission from [\ion{O}{4}] at 25.9\micron.}
\label{excessOIV}
\end{figure}

Figure \ref{excessOIV} shows the IRS spectra of all three galaxies (IDs 6, 50, and 55).  Both galaxies IDs 6 and 50 have spectra similar to the IRAGN composite spectrum (figure \ref{composite spec}) and the spectra of the other IRAGN (figure \ref{decomps}), in our sample.  In contrast, galaxy ID 55 has an IRS spectrum more characteristic of the other non-IRAGN (see figures \ref{decomps} and \ref{composite spec}).

\begin{figure}
\epsscale{1.2}
\plotone{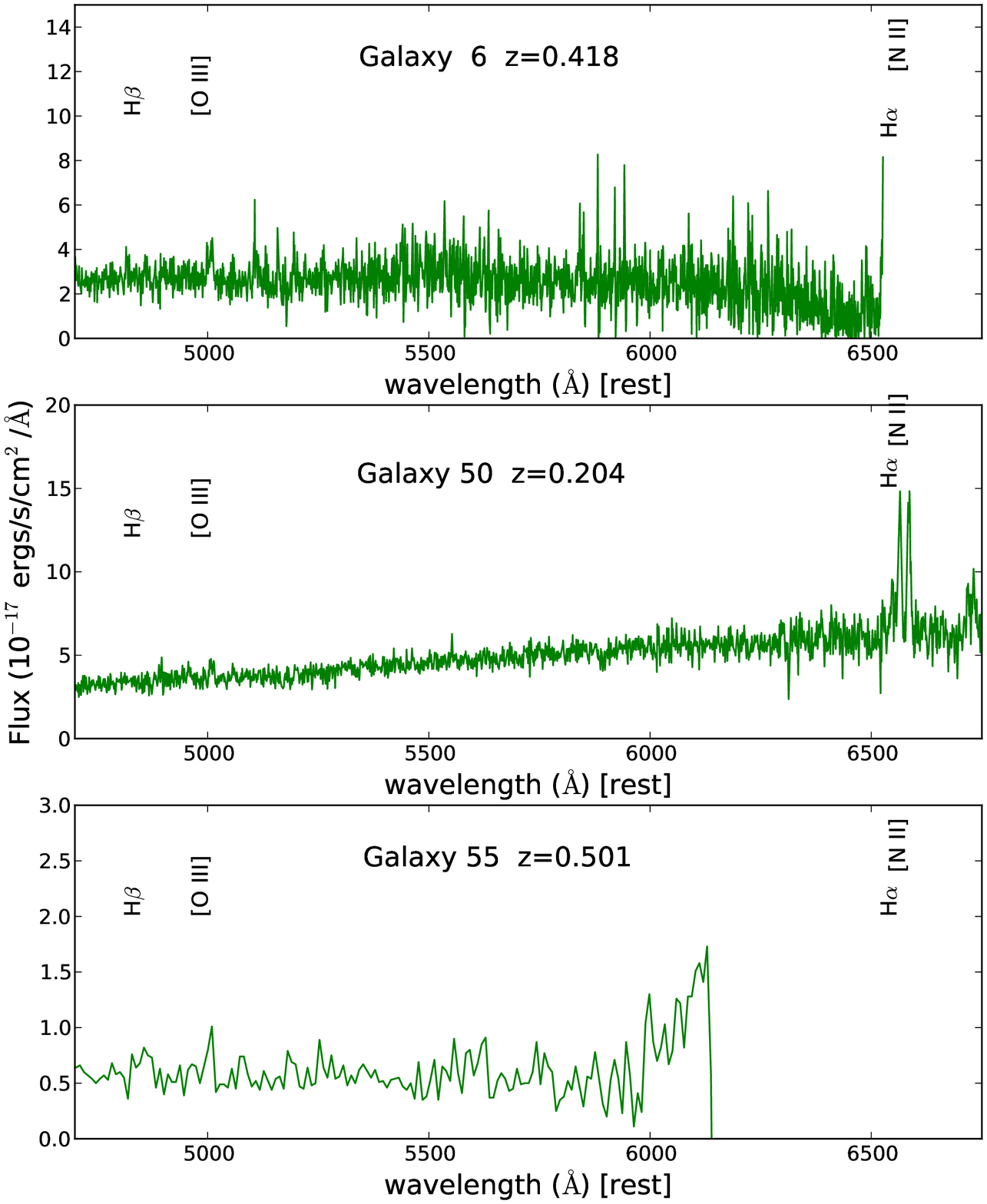}
\caption{The optical spectra of three non-IRAGN galaxies in our IRS sample with excess [\ion{O}{4}] emission and no other indications of an AGN.  Galaxy IDs 6 and 50 are optical spectra from AGES.  Galaxy 55 is from the FLS and has an optical spectrum from \citet{Papovich2006}.  The emission lines of [\ion{O}{3}] $\lambda$5007$\mathrm{\AA}$, H$\beta$ $\lambda$4861$\mathrm{\AA}$, [\ion{N}{2}] $\lambda$6583$\mathrm{\AA}$, and H$\alpha$ $\lambda$6563$\mathrm{\AA}$ have been marked to indicate the atomic lines used for optical classification of galaxies as star-forming, composite, or AGN \citep{Kewley2001, Kauffmann2003}.  The redshifts listed for each galaxy are the given optical redshifts from AGES (see table \ref{table galaxy stats}).}
\label{excessOIV opt}
\end{figure}

We looked for other indications of obscured AGN in the optical
spectroscopy (see figure \ref{excessOIV opt}) of galaxy IDs 6 and 50, both of which have optical
spectra from AGES (galaxy ID 55 lies in the FLS and has optical spectra from \citealt{Papovich2006}, but the spectral coverage does not extend to H$\alpha$ and [\ion{N}{2}]).  We show the portion of the optical spectrum covering the emission lines of [\ion{O}{3}] $\lambda$5007$\mathrm{\AA}$, H$\beta$ $\lambda$4861$\mathrm{\AA}$, [\ion{N}{2}] $\lambda$6583$\mathrm{\AA}$, and H$\alpha$ $\lambda$6563$\mathrm{\AA}$.  The optical spectra of all three galaxies (IDs 6, 50, and 55) are typical of star-forming galaxies, and therefore the AGN in these galaxies must be very obscured.  Using an optical BPT classification \citep{Kewley2001, Kauffmann2003} of the galaxies, the galaxies would likely be classified as either star-forming or composite sources.  Galaxy ID 50 has all four emission lines used for optical classification and is the most likely source showing any indication of an AGN being present in its optical spectrum.

Of the three non-IRAGN with unusually high [\ion{O}{4}] emission (figure \ref{Ne II O IV}, bottom panel), galaxy IDs 6 and 50 both have low L$_{\mathrm{PAH}}$/L$_{\mathrm{LIR}}$ ratios, and are quite similar to the values observed in the IRAGN.  Galaxy ID 55 has one of the highest L$_{\mathrm{PAH}}$/L$_{\mathrm{LIR}}$ ratios in the entire IRS sample.  As expected, galaxy IDs 6 and 50 both have low [Ne II] luminosities similar to the five IRAGN with high [\ion{O}{4}] emission (and most of the IRAGN, shown in figure \ref{Ne II O IV}).  Further evidence for galaxy IDs 6 and 50 being obscured AGN may be seen in that they have optical colors and magnitudes located in the ``green valley'', similar to the other IRAGN.  Furthermore, galaxy No. 55 is about the same as the median value of the non-IRAGN subsample's [Ne II] luminosity (refer to table \ref{table PAH emission fits} for fitted flux values).  Therefore, galaxy ID 55 is clearly a candidate for an IR-luminous galaxy with a deeply obscured AGN that is otherwise undetected.  Further studies of these galaxies would allow for tests of galaxies with deeply obscured AGN, including multi-wavelength observations.

\section{Conclusions}
\label{conclusions}

We have studied a sample of 65 galaxies with spectroscopic redshifts of 0.02 $<$ z $<$ 0.6 using \spitzer\ IRS mid-IR spectroscopy, combined with \spitzer\ imaging and ground-based observations that have flux densities of $f_{\nu}$(24\micron) $>$ 1.2 mJy.  The galaxies have a total IR luminosity from $\sim 10^{10}$L$_{\odot}$ to more than $10^{12}$L$_{\odot}$, with a median of 3.0$\times 10^{11}$L$_{\odot}$, which we estimate using the observed MIPS 24 \micron\ data (and 70 and 160~\micron\ when available) combined with \citet{Rieke2009} IR SED templates.

We use an IRAC color--color selection \citep{Stern2005} to identify IRAGN and non-IRAGN for galaxies in our sample.   Our sample includes a wide range of IR-luminous galaxies at redshifts where such galaxies dominate the cosmic IR-luminosity density and SFR-density and they therefore provide insight into the physical processes within these galaxies. 

We have shown the measured PAH emission can contribute 11$\%$ (75th percentile) or more of the total IR luminosity and half that or slightly more can come from the 7.7$\micron$ feature alone, for either the non-IRAGN or IRAGN.  The detection frequency of PAH emission (for all features) in the non-IRAGN sample is uniformly more than twice that for IRAGN, except for the
11.3$\micron$ feature which is detected only about 1.5 times more frequently in non-IRAGN.

We do not see a significant difference between IRAGN galaxies and non-IRAGN galaxies for short-to-long PAH ratios (namely the 7.7$\micron$-to-11.3$\micron$ ratio versus the [\ion{Ne}{3}]/[\ion{Ne}{2}] ratio) in contrast to other studies \citep{Smith2007, ODowd2009, Wu2010}.  However, the sample galaxies used by these other studies include more nearby galaxies, where the IRS slit encompasses only the nuclear emission.  In our sample, the IRS slit contains more of the integrated emission from the galaxy.  The data used by these other studies are therefore more sensitive to the destruction of smaller grains from AGN-driven shocks that lower the emission of shorter wavelength PAH features \citep[6.2$\micron$ and 7.7$\micron$ features,][]{Diamond-Stanic2010}.  We investigated this by using the H$_{\mathrm{2}}$ S(3) rotational line (at 9.67$\micron$) and found that none of our galaxies had strong H$_{\mathrm{2}}$ emission.  In fact, we did not have many detections of H$_{\mathrm{2}}$ emission (only 10 of 65 galaxies have H$_{\mathrm{2}}$ flux detections at $\geq$ 3$\sigma$) for our sample.  This result suggests that if shocks are present in the members of our sample they are not strong enough to dramatically alter the observed short-to-long PAH ratios.

The PAH emission correlates very tightly with the [\ion{Ne}{2}] emission.  We conclude that both processes correlate with the total SFR, implying that we can use the PAH emission to estimate the contribution of star-formation to the total IR luminosity in the IRAGN.  As expected for the IRAGN, the dominant source of IR luminosity comes from the AGN.  In fact, we find only about 10$\%$--50$\%$ of the contribution comes from star-formation for most of the IRAGN in our sample.  The IRAGN have lower [\ion{Ne}{2}] emission indicating that the presence of an AGN does not require strong star-formation, and/or that the presence of an AGN suppresses star-formation.  The PAH emission does not correlate as well with [\ion{O}{4}] emission, which is more sensitive to the ionization radiation from the AGN.  At a basic level, this shows that AGN do not increase PAH emission.  We do identify a population of galaxies with low PAH emission and excess [\ion{O}{4}] emission, indicative of heavily-obscured AGN.   Of the eight galaxies with excess [\ion{O}{4}] emission, three are non-IRAGN, with no other indications of AGN from their IRAC colors or optical spectroscopy.  Nonetheless, two of these three have suppressed PAH features indicating they have obscured AGN; the third shows no such effects, and its AGN may be very strongly obscured.

Investigating the color-magnitude relation, the IRAGN span the full range of $(u-r)_{0.1}$ colors as do the non-IRAGN for the full sample of IR-luminous galaxies.  The optical colors of the IRAGN are more consistent with the distribution of other AGN samples, mostly populating the top of the star-forming blue sequence to the red sequence \citep[see, e.g.,][]{Nandra2007, Weiner2007}.  Because no significant difference is seen between IRAGN and non-IRAGN, it is unclear if the presence of an AGN is driving redder $(u-r)_{0.1}$ colors in the IRAGN of IR luminous galaxies.  Moreover, most of the IRAGN have lower star-formation (because they show weak [\ion{Ne}{2}] emission).  Therefore, star-formation is not a necessary component for an AGN to persist.

\acknowledgements 

We thank the anonymous referee for valuable comments that improved the quality of this work.  We thank our colleagues on the NDWFS, AGES teams.  We thank JD Smith for comments that helped improve the manuscript.  This work made use of images and/or data products provided by the NOAO Deep Wide-Field Survey \citep{Jannuzi1999, Jannuzi2004, Dey2004}, which is supported by the National Optical Astronomy Observatory (NOAO). NOAO is operated by AURA, Inc., under a cooperative agreement with the National Science Foundation.  Support for this work was provided to the authors by the George P. and Cynthia Woods Mitchell Institute for Fundamental Physics and Astronomy.  The research of AD is supported by NOAO, which is operated by the Association of Universities for Research in Astronomy, Inc., under a cooperative agreement with the NSF.  This work is based  in part on observations and archival data obtained with the Spitzer Space Telescope, which is operated by the Jet Propulsion Laboratory, California Institute of Technology under a contract with NASA.  Partial support for this work was provided by NASA through awards 1255094 and 1365085 issued by JPL/Caltech.

\bibliography{myrefs}{}


\begin{deluxetable}{ccccccrrrccc}
\tablecaption{Galaxy Classifications and Locations\label{table galaxy stats}}
\tablecolumns{12}
\tabletypesize{\scriptsize}
\tablewidth{0pc}
\tablehead{
\colhead{ID} & \colhead{AOR} & \colhead{R.A.} & \colhead{Dec.} & \colhead{z(opt)} & \colhead{z(IRS)} & \colhead{f(24$\micron$)} & \colhead{f(70$\micron$)} & \colhead{f(160$\micron$)} & \colhead{L(IR)$\mathrm{_{R}}$} & \colhead{AGN} \\
\colhead{} & \colhead{} & \colhead{} & \colhead{} & \colhead{} & \colhead{} & \colhead{[mJy]} & \colhead{[mJy]} & \colhead{[mJy]} & \colhead{[$10^{10}$L$_\odot$]} & \colhead{}
}
\startdata
1 & 14127872 & 14h25m00.18s & +32d59m50.0s & 0.296 & 0.304 & 4.97 $\pm$ 0.02 & 60.7 $\pm$ 1.09 & ... & 31.9 & \\ 
2 & 21757184 & 14h25m27.91s & +33d47m47.0s & 0.509 & 0.516 & 5.38 $\pm$ 0.02 & 17.4 $\pm$ 1.38 & ... & 94.6 & Y \\ 
3 & 14129920 & 14h25m44.96s & +33d37m05.0s & 0.321 & 0.344 & 3.01 $\pm$ 0.01 & ... & ... & 24.8 & Y \\ 
4 & 14087680 & 14h25m52.68s & +34d02m40.1s & 0.562 & 0.565 & 11.36 $\pm$ 0.02 & 258.2 $\pm$ 1.03 & 413.0 $\pm$ 1.12 & 268. & Y \\ 
5 & 14133248 & 14h26m07.93s & +34d50m45.0s & 0.405 & 0.407 & 2.77 $\pm$ 0.01 & 47.5 $\pm$ 1.09 & 143.9 $\pm$ 1.21 & 39.0 & \\ 
6 & 14133504 & 14h26m18.65s & +34d57m03.0s & 0.418 & 0.415 & 1.58 $\pm$ 0.01 & ... & ... & 35.8 & \\ 
7 & 14084352 & 14h26m23.88s & +32d44m35.8s & 0.171 & 0.173 & 12.17 $\pm$ 0.02 & 304.1 $\pm$ 1.03 & 474.2 $\pm$ 1.04 & 31.1 & Y \\ 
8 & 14126080 & 14h26m59.13s & +33d33m05.0s & 0.150 & 0.150 & 16.12 $\pm$ 0.19 & 178.6 $\pm$ 1.25 & ... & 22.9 & \\ 
9 & 21758464 & 14h27m44.37s & +34d29m08.2s & 0.365 & 0.365 & 1.56 $\pm$ 0.01 & 109.4 $\pm$ 1.06 & 234.4 $\pm$ 1.12 & 59.1 & \\ 
10 & 21758208 & 14h27m50.32s & +35d04m51.2s & 0.211 & 0.210 & 4.82 $\pm$ 0.01 & 87.1 $\pm$ 1.07 & 206.1 $\pm$ 1.15 & 17.2 & \\ 
11 & 14132480 & 14h28m19.51s & +33d51m50.0s & 0.488 & 0.490 & 1.70 $\pm$ 0.01 & 22.3 $\pm$ 1.24 & 101.9 $\pm$ 1.24 & 53.4 & \\ 
12 & 14132992 & 14h28m19.57s & +34d23m34.0s & 0.474 & 0.493 & 2.05 $\pm$ 0.01 & 18.9 $\pm$ 1.22 & ... & 54.0 & Y \\ 
13 & 14086144 & 14h28m49.79s & +34d32m40.2s & 0.216 & 0.218 & 8.09 $\pm$ 0.03 & 253.5 $\pm$ 1.04 & 598.4 $\pm$ 1.06 & 47.5 & \\ 
14 & 14132736 & 14h29m19.77s & +34d15m06.0s & 0.424 & 0.425 & 1.64 $\pm$ 0.01 & 23.3 $\pm$ 1.16 & 92.7 $\pm$ 1.26 & 37.9 & \\ 
15 & 14132224 & 14h29m35.97s & +33d37m13.0s & 0.419 & 0.422 & 1.90 $\pm$ 0.01 & 33.3 $\pm$ 1.17 & 131.8 $\pm$ 1.20 & 38.8 & \\ 
16 & 21758720 & 14h29m50.66s & +35d08m42.7s & 0.366 & 0.380 & 5.87 $\pm$ 0.01 & 19.2 $\pm$ 1.25 & ... & 53.8 & Y \\ 
17 & 21755648 & 14h29m51.38s & +32d50m36.0s & 0.270 & 0.268 & 3.39 $\pm$ 0.01 & 19.1 $\pm$ 1.22 & ... & 16.8 & Y \\ 
18 & 14129408 & 14h30m00.39s & +35d38m14.0s & 0.240 & 0.240 & 9.11 $\pm$ 0.02 & 125.3 $\pm$ 1.06 & 146.9 $\pm$ 1.15 & 28.0 & \\ 
19 & 21756672 & 14h30m19.82s & +33d40m47.2s & 0.231 & 0.233 & 2.62 $\pm$ 0.01 & 44.9 $\pm$ 1.15 & 97.3 $\pm$ 1.21 & 11.4 & \\ 
20 & 14082304 & 14h30m24.46s & +32d56m16.4s & 0.042 & 0.042 & 36.68 $\pm$ 0.15 & 483.1 $\pm$ 1.04 & 824.1 $\pm$ 1.08 & 4.56 & \\ 
21 & 14086400 & 14h31m14.77s & +33d46m23.0s & 0.230 & 0.231 & 4.53 $\pm$ 0.03 & 153.5 $\pm$ 1.09 & 347.5 $\pm$ 1.07 & 31.1 & \\ 
22 & 14081792 & 14h31m19.76s & +35d34m18.0s & 0.034 & 0.036 & 33.58 $\pm$ 0.11 & 537.0 $\pm$ 1.03 & 691.8 $\pm$ 1.06 & 3.11 & \\ 
23 & 14082048 & 14h31m21.12s & +35d37m21.8s & 0.035 & 0.036 & 33.18 $\pm$ 0.46 & 1318.3 $\pm$ 1.04 & 2779.7 $\pm$ 1.04 & 11.4 & \\ 
24 & 14130432 & 14h31m21.88s & +34d40m46.0s & 0.348 & 0.354 & 3.19 $\pm$ 0.01 & 17.1 $\pm$ 1.20 & ... & 27.4 & Y \\ 
25 & 14080768 & 14h31m25.43s & +33d13m49.7s & 0.022 & 0.022 & 33.98 $\pm$ 0.54 & 794.3 $\pm$ 1.06 & 1011.6 $\pm$ 1.07 & 2.04 & \\ 
26 & 21756160 & 14h31m33.98s & +33d45m16.0s & 0.490 & 0.494 & 2.46 $\pm$ 0.01 & 22.8 $\pm$ 1.22 & ... & 54.4 & Y \\ 
27 & 14081536 & 14h31m56.23s & +33d38m33.1s & 0.034 & 0.035 & 29.27 $\pm$ 0.50 & ... & ... & 1.67 & \\ 
28 & 14083840 & 14h32m28.36s & +34d58m38.8s & 0.129 & 0.130 & 3.15 $\pm$ 0.02 & 137.1 $\pm$ 1.13 & 258.8 $\pm$ 1.22 & 7.63 & \\ 
29 & 14086912 & 14h32m34.90s & +33d28m32.3s & 0.249 & 0.253 & 3.85 $\pm$ 0.02 & 338.8 $\pm$ 1.03 & 395.4 $\pm$ 1.09 & 46.6 & \\ 
30 & 14086656 & 14h32m39.56s & +35d01m51.3s & 0.236 & 0.237 & 10.73 $\pm$ 0.02 & 236.6 $\pm$ 1.03 & 349.9 $\pm$ 1.07 & 42.9 & \\ 
31 & 14131968 & 14h32m52.49s & +33d11m53.0s & 0.401 & 0.395 & 2.58 $\pm$ 0.01 & 28.2 $\pm$ 1.21 & ... & 32.4 & \\ 
32 & 14128128 & 14h33m26.18s & +33d05m58.0s & 0.243 & 0.245 & 4.28 $\pm$ 0.02 & 62.1 $\pm$ 1.08 & 119.4 $\pm$ 1.21 & 16.6 & \\ 
33 & 21755904 & 14h33m33.34s & +33d09m22.0s & 0.353 & 0.356 & 2.37 $\pm$ 0.01 & ... & ... & 24.9 & Y \\ 
34 & 21757440 & 14h33m42.17s & +34d56m51.0s & 0.494 & 0.494 & 2.24 $\pm$ 0.01 & ... & ... & 54.4 & \\ 
35 & 14083584 & 14h34m45.32s & +33d13m46.1s & 0.073 & 0.075 & 14.46 $\pm$ 0.03 & 288.4 $\pm$ 1.03 & 485.3 $\pm$ 1.08 & 7.01 & \\ 
36 & 14130176 & 14h34m53.85s & +34d27m44.0s & 0.329 & 0.324 & 3.56 $\pm$ 0.01 & 29.4 $\pm$ 1.15 & ... & 25.5 & \\ 
37 & 21756416 & 14h35m00.65s & +33d29m23.0s & 0.274 & 0.285 & 2.61 $\pm$ 0.01 & 70.0 $\pm$ 1.08 & 140.6 $\pm$ 1.16 & 20.1 & \\ 
38 & 14081024 & 14h35m18.21s & +35d07m08.3s & ... & 0.029 & 169.23 $\pm$ 3.72 & 3926.4 $\pm$ 1.04 & 5571.9 $\pm$ 1.04 & 19.7 & \\ 
39 & 14130688 & 14h35m19.42s & +35d36m22.0s & 0.316 & 0.318 & 2.92 $\pm$ 0.01 & 26.5 $\pm$ 1.24 & ... & 20.8 & Y \\ 
40 & 21758976 & 14h35m35.47s & +33d25m44.4s & 0.244 & 0.251 & 6.07 $\pm$ 0.02 & 104.2 $\pm$ 1.06 & 145.5 $\pm$ 1.18 & 24.9 & \\ 
41 & 14134272 & 14h36m06.84s & +35d09m27.0s & 0.525 & 0.525 & 1.22 $\pm$ 0.01 & 21.0 $\pm$ 1.22 & ... & 63.1 & \\ 
42 & 14125824 & 14h36m19.14s & +33d29m17.0s & 0.188 & 0.190 & 7.07 $\pm$ 0.05 & ... & ... & 17.8 & \\ 
43 & 14087168 & 14h36m28.12s & +33d33m58.0s & 0.265 & 0.269 & 10.82 $\pm$ 0.03 & 223.4 $\pm$ 1.04 & 267.3 $\pm$ 1.11 & 51.0 & \\ 
44 & 14128640 & 14h36m33.16s & +33d48m05.0s & 0.218 & 0.245 & 6.43 $\pm$ 0.02 & 12.4 $\pm$ 1.38 & ... & 25.4 & Y \\ 
45 & 14129152 & 14h36m36.65s & +34d50m34.0s & 0.279 & 0.274 & 4.05 $\pm$ 0.02 & 39.2 $\pm$ 1.12 & ... & 21.0 & \\ 
46 & 14081280 & 14h36m41.23s & +34d58m24.2s & 0.030 & 0.030 & 16.75 $\pm$ 0.25 & ... & ... & 0.84 & \\ 
47 & 14130944 & 14h37m11.26s & +35d40m36.0s & 0.362 & 0.362 & 3.42 $\pm$ 0.02 & 223.9 $\pm$ 1.04 & 236.0 $\pm$ 1.16 & 64.0 & \\ 
48 & 21757952 & 14h37m23.73s & +35d07m35.0s & 0.579 & 0.551 & 4.78 $\pm$ 0.01 & 30.5 $\pm$ 1.15 & ... & 96.2 & \\ 
49 & 21757696 & 14h37m52.92s & +35d32m51.4s & 0.563 & 0.565 & 2.36 $\pm$ 0.01 & 68.7 $\pm$ 1.10 & 217.3 $\pm$ 1.10 & 101. & \\ 
50 & 21756928 & 14h38m09.88s & +35d27m37.8s & 0.204 & 0.201 & 2.70 $\pm$ 0.02 & 31.7 $\pm$ 1.15 & ... & 8.74 & \\ 
51 & 14127104 & 17h12m39.64s & +58d41m48.0s & 0.165 & 0.166 & 10.06 $\pm$ 0.13 & ... & ... & 18.3 & \\ 
52 & 14131712 & 17h13m08.57s & +60d16m21.0s & 0.332 & 0.333 & 4.65 $\pm$ 0.09 & 76.2 $\pm$ 12.20 & ... & 36.7 & \\ 
53 & 14135552 & 17h14m27.02s & +58d38m36.0s & 0.562 & 0.565 & 1.30 $\pm$ 0.06 & ... & ... & 75.6 & \\ 
54 & 14127360 & 17h14m37.44s & +59d56m48.1s & 0.196 & 0.198 & 12.38 $\pm$ 0.14 & ... & ... & 30.8 & \\ 
55 & 14135808 & 17h15m25.74s & +60d04m24.0s & 0.501 & 0.499 & 1.34 $\pm$ 0.06 & ... & ... & 55.8 & \\ 
56 & 14127616 & 17h15m42.00s & +59d16m57.4s & 0.116 & 0.118 & 26.78 $\pm$ 0.21 & ... & ... & 19.7 & \\ 
57 & 14126336 & 17h19m16.60s & +59d34m49.0s & 0.166 & 0.170 & 17.06 $\pm$ 0.17 & 177.5 $\pm$ 27.00 & ... & 30.2 & \\ 
58 & 14131200 & 17h20m25.19s & +59d15m03.0s & 0.305 & 0.307 & 3.28 $\pm$ 0.08 & ... & ... & 21.3 & \\ 
59 & 14134528 & 17h20m46.74s & +59d22m23.0s & 0.539 & 0.526 & 1.26 $\pm$ 0.06 & ... & ... & 63.4 & \\ 
60 & 14134784 & 17h21m17.75s & +58d51m20.0s & 0.512 & 0.493 & 1.54 $\pm$ 0.06 & ... & ... & 54.1 & Y \\ 
61 & 14135040 & 17h21m18.31s & +58d46m01.0s & 0.555 & 0.559 & 1.22 $\pm$ 0.05 & ... & ... & 73.5 & Y \\ 
62 & 14126592 & 17h22m20.27s & +59d09m49.0s & 0.179 & 0.179 & 9.66 $\pm$ 0.13 & 180.4 $\pm$ 27.50 & ... & 21.3 & \\ 
63 & 14131456 & 17h23m48.13s & +59d01m54.0s & 0.321 & 0.324 & 4.26 $\pm$ 0.09 & ... & ... & 29.9 & \\ 
64 & 14126848 & 17h24m00.61s & +59d02m28.0s & 0.178 & 0.180 & 12.82 $\pm$ 0.14 & ... & ... & 26.5 & \\ 
65 & 14135296 & 17h25m03.37s & +59d11m09.0s & 0.514 & 0.515 & 1.30 $\pm$ 0.06 & ... & ... & 60.2 & \\
\enddata
\tablecomments{Optical redshifts [z(opt)] are from the AGES catalog \citep{Kochanek2012} and \citet{Papovich2006} for FLS sources.  IRS redshifts [z(IRS)] are from the fits as discussed in section \ref{fits}. The 24\micron\ flux densities are from the AGES catalog \citep{Kochanek2012} and \citet{Papovich2006} for FLS sources and discussed in sections \ref{AGES data} and \ref{FLS data}.  The 70\micron\ and 160\micron\ flux densities are from MAGES (B. Jannuzi et al., in prep) and discussed in sections \ref{AGES data} and \ref{FLS data}.  L$_{\mathrm{IR}}$ values are described in section \ref{LIR} and appendix \ref{MAGES LIR} and the selection of IRAGN in section \ref{AGN selection}.}
\end{deluxetable}

\begin{deluxetable}{ccccccc}
\tablecaption{Fitted Fluxes of the Most Prominent PAH Emission Features\label{table PAH emission fits}}
\tablecolumns{7}
\tabletypesize{\footnotesize}
\tablewidth{0pc}
\tablehead{
\colhead{ID} & \colhead{F$_{6.2}$} & \colhead{F$_{7.7}$} & \colhead{F$_{8.6}$} & \colhead{F$_{11.3}$} & \colhead{F$_{12.7}$} & \colhead{F$_{17.0}$}\\ 
\colhead{} & \colhead{[$10^{-17}$]} & \colhead{[$10^{-16}$]} & \colhead{[$10^{-17}$]} & \colhead{[$10^{-17}$]} & \colhead{[$10^{-17}$]} & \colhead{[$10^{-17}$]}
}
\startdata
 1 & 4.50 $\pm$ 1.45 & 2.48 $\pm$ 0.41 & 19.3 $\pm$ 2.30 & 0.39 $\pm$ 0.54 & 6.42 $\pm$ 0.56 & $...$\\ 
 2 & 0.30 $\pm$ 1.47 & 0.67 $\pm$ 0.72 & 0.79 $\pm$ 2.67 & 1.07 $\pm$ 0.75 & 1.48 $\pm$ 1.14 & 1.73 $\pm$ 0.54\\ 
 3 & $...$ & 0.26 $\pm$ 0.22 & 1.09 $\pm$ 0.75 & 0.39 $\pm$ 0.15 & 0.23 $\pm$ 0.33 & 0.99 $\pm$ 1.55\\ 
 4 & 7. 4 $\pm$ 0.69 & 3.31 $\pm$ 0.34 & 12.2 $\pm$ 1.46 & 6.97 $\pm$ 1.74 & 6.65 $\pm$ 2.62 & 2.57 $\pm$ 0.85\\ 
 5 & 6.37 $\pm$ 0.60 & 2.12 $\pm$ 0.24 & 6.92 $\pm$ 0.98 & 6.49 $\pm$ 0.77 & 3.49 $\pm$ 0.68 & 3.19 $\pm$ 2.28\\ 
 6 & 0.58 $\pm$ 0.48 & 0.30 $\pm$ 0.17 & 0.78 $\pm$ 0.41 & 1.02 $\pm$ 0.19 & 0.12 $\pm$ 0.29 & 0.74 $\pm$ 0.90\\ 
 7 &  7.5 $\pm$ 1.62 & 5.58 $\pm$ 0.48 & 14.6 $\pm$ 1.57 & 13.5 $\pm$ 1.49 & 7.30 $\pm$ 2.49 & 5.12 $\pm$ 3.88\\ 
 8 & 17.2 $\pm$ 0.94 & 6.45 $\pm$ 0.33 &  8.0 $\pm$ 1.03 & 16.6 $\pm$ 0.90 & 9.28 $\pm$ 1.48 & 8.38 $\pm$ 1.56\\ 
 9 & 11.4 $\pm$ 1.52 & 3.46 $\pm$ 0.67 & 12.4 $\pm$ 2.39 & 9.90 $\pm$ 1.85 & 6.50 $\pm$ 1.16 & 2.11 $\pm$ 0.66\\ 
10 & 7.48 $\pm$ 5.79 & 2.34 $\pm$ 0.79 & 7.76 $\pm$ 2.03 & 7.95 $\pm$ 2.16 & 7.84 $\pm$ 2.58 & 6.69 $\pm$ 3.23\\ 
11 & 7.01 $\pm$ 0.68 & 2.34 $\pm$ 0.23 & 7.49 $\pm$ 1.23 & 6.61 $\pm$ 0.82 & 4.50 $\pm$ 0.51 & 2.14 $\pm$ 1.79\\ 
12 & 0.88 $\pm$ 0.54 & 0.56 $\pm$ 0.23 & 1.27 $\pm$ 0.78 & 0.70 $\pm$ 0.40 & 0.46 $\pm$ 0.17 & 0.11 $\pm$ 0.27\\ 
13 & 26.7 $\pm$ 1.01 & 9.80 $\pm$ 0.51 & 29.2 $\pm$ 1.68 & 23.8 $\pm$ 1.75 & 14.8 $\pm$ 1.53 & 8.19 $\pm$ 3.29\\ 
14 & 4.45 $\pm$ 0.68 & 1.31 $\pm$ 0.27 & 4.98 $\pm$ 1.37 & 7.59 $\pm$ 1.20 & 1.63 $\pm$ 0.87 & 2.62 $\pm$ 2.14\\ 
15 & 5.12 $\pm$ 0.70 & 1.75 $\pm$ 0.28 & 5.69 $\pm$ 1.09 & 4.95 $\pm$ 0.77 & 3.76 $\pm$ 0.71 & 2.26 $\pm$ 1.94\\ 
16 & 1.42 $\pm$ 1.69 & 0.16 $\pm$ 0.74 & 0.95 $\pm$ 2.01 & 2.15 $\pm$ 1.67 & 1.80 $\pm$ 1.72 & 1.28 $\pm$ 0.85\\ 
17 & 2.07 $\pm$ 3.19 & 0.59 $\pm$ 0.54 & 2.28 $\pm$ 2.02 & 2.10 $\pm$ 1.13 & 1.14 $\pm$ 1.42 & 1.73 $\pm$ 2.26\\ 
18 & 9.08 $\pm$ 0.91 & 2.60 $\pm$ 0.34 & 10.1 $\pm$ 1.05 & 8.75 $\pm$ 0.85 & 3.96 $\pm$ 0.71 & 6.74 $\pm$ 1.14\\ 
19 & 1.76 $\pm$ 1.61 & 0.76 $\pm$ 0.43 & 2.23 $\pm$ 1.31 & 2.44 $\pm$ 1.12 & 1.55 $\pm$ 1.14 & 0.91 $\pm$ 0.60\\ 
20 & 28.6 $\pm$ 0.77 & 9.29 $\pm$ 0.37 & 28.8 $\pm$ 1.08 & 26.3 $\pm$ 0.86 & 16.1 $\pm$ 0.91 & 18.1 $\pm$ 1.51\\ 
21 & 10.8 $\pm$ 0.69 & 3.88 $\pm$ 0.33 & 21.4 $\pm$ 1.22 & 10.9 $\pm$ 1.24 & 5.57 $\pm$ 1.18 & 2.99 $\pm$ 2.42\\ 
22 & 43.1 $\pm$ 0.87 & 15.7 $\pm$ 0.36 &223.1 $\pm$ 1.08 & 42.6 $\pm$ 0.76 & 24.9 $\pm$ 1.14 & 30.0 $\pm$ 1.75\\ 
23 & 81.0 $\pm$ 1.51 & 30.9 $\pm$ 0.49 & 79.6 $\pm$ 1.82 &231.3 $\pm$ 1.39 & 38.4 $\pm$ 1.32 & 58.1 $\pm$ 2.25\\ 
24 & 1.18 $\pm$ 0.71 & 0.60 $\pm$ 0.28 & 2.40 $\pm$ 0.98 & 1.39 $\pm$ 0.40 & 0.47 $\pm$ 0.16 & 0.74 $\pm$ 1.87\\ 
25 & 22.1 $\pm$ 0.90 & 7.34 $\pm$ 0.40 & 21.8 $\pm$ 0.87 & 21.8 $\pm$ 0.64 & 9.89 $\pm$ 0.89 &256.8 $\pm$25.17\\ 
26 & 3.43 $\pm$ 1.00 & 1.22 $\pm$ 0.31 & 3.85 $\pm$ 1.00 & 3.43 $\pm$ 0.66 & 2.65 $\pm$ 1.16 & 1.96 $\pm$ 1.50\\ 
27 & 27.2 $\pm$ 0.88 & 10.7 $\pm$ 0.35 & 37.7 $\pm$ 1.10 & 30.1 $\pm$ 0.86 & 16.1 $\pm$ 1.06 &270.9 $\pm$ 1.76\\ 
28 & 7.17 $\pm$ 0.67 & 2.65 $\pm$ 0.22 & 7.75 $\pm$ 0.66 & 7.86 $\pm$ 0.60 & 5.62 $\pm$ 0.59 & 5.78 $\pm$ 1.02\\ 
29 & 5.88 $\pm$ 0.51 & 2.60 $\pm$ 0.19 & 8.07 $\pm$ 0.64 & 6.09 $\pm$ 0.69 & 2.15 $\pm$ 0.39 & 3.41 $\pm$ 1.03\\ 
30 & 11.9 $\pm$ 0.76 & 4.33 $\pm$ 0.41 & 16.1 $\pm$ 1.25 & 14.4 $\pm$ 1.24 & 8.55 $\pm$ 0.98 & 5.06 $\pm$ 2.62\\ 
31 & 1.81 $\pm$ 0.66 & 0.54 $\pm$ 0.12 & 1.84 $\pm$ 0.78 & 3.01 $\pm$ 0.49 & 0.83 $\pm$ 0.55 & 3.19 $\pm$ 1.54\\ 
32 & 5.40 $\pm$ 0.91 & 1.47 $\pm$ 0.40 & 5.83 $\pm$ 1.08 & 6.79 $\pm$ 0.80 & 2.90 $\pm$ 0.60 & 2.60 $\pm$ 0.95\\ 
33 & 0.89 $\pm$ 0.52 & 0.32 $\pm$ 0.08 & 1.23 $\pm$ 0.66 & 0.82 $\pm$ 0.52 & 0.76 $\pm$ 0.84 & 1.11 $\pm$ 0.66\\ 
34 & 0.55 $\pm$ 0.47 & 0.31 $\pm$ 0.18 & 1.50 $\pm$ 0.85 & 0.45 $\pm$ 0.67 & 0.84 $\pm$ 1.20 & 0.25 $\pm$ 0.27\\ 
35 & 10.6 $\pm$ 1.06 & 3.84 $\pm$ 0.40 & 12.2 $\pm$ 1.22 & 13.7 $\pm$ 1.00 & 7.25 $\pm$ 1.01 & 12.7 $\pm$ 2.08\\ 
36 & 1.61 $\pm$ 1.13 & 0.63 $\pm$ 0.17 & 1.88 $\pm$ 0.82 & 2.08 $\pm$ 0.52 & 1.49 $\pm$ 0.51 & 0.37 $\pm$ 0.53\\ 
37 & 1.96 $\pm$ 0.68 & 1.07 $\pm$ 0.16 & 2.83 $\pm$ 0.67 & 3.24 $\pm$ 0.88 & 2.85 $\pm$ 0.84 & 0.27 $\pm$ 0.33\\ 
38 & 61.5 $\pm$ 0.89 &385.5 $\pm$ 0.43 & 66.5 $\pm$ 1.22 & 77.2 $\pm$ 1.04 & 39.6 $\pm$ 1.18 & 173. $\pm$ 4.74\\ 
39 & 1.67 $\pm$ 1.03 & 0.79 $\pm$ 0.29 & 2.71 $\pm$ 1.16 & 1.74 $\pm$ 0.64 & 1.28 $\pm$ 0.37 & 2.01 $\pm$ 2.61\\ 
40 & 7.97 $\pm$ 3.22 & 2.41 $\pm$ 0.71 & 7.85 $\pm$ 2.82 & 6.70 $\pm$ 2.50 & 5.25 $\pm$ 2.36 & 5.57 $\pm$ 8.23\\ 
41 & 1.67 $\pm$ 0.42 & 0.60 $\pm$ 0.19 & 2.06 $\pm$ 0.88 & 3.71 $\pm$ 0.43 & 2.56 $\pm$ 0.73 & 2.24 $\pm$ 1.75\\ 
42 & 7.96 $\pm$ 1.22 & 2.37 $\pm$ 0.27 & 7.35 $\pm$ 1.15 & 4.79 $\pm$ 1.12 & 9.65 $\pm$ 0.37 & 7.60 $\pm$ 1.54\\ 
43 & 10.1 $\pm$ 0.96 & 4.13 $\pm$ 0.35 & 10.7 $\pm$ 1.14 & 7.72 $\pm$ 1.43 & 5.83 $\pm$ 1.64 & 4.68 $\pm$ 2.36\\ 
44 & 0.80 $\pm$ 0.84 & 0.68 $\pm$ 0.34 & 0.55 $\pm$ 0.83 & 2.19 $\pm$ 0.39 & 0.24 $\pm$ 0.13 & 0.01 $\pm$ 0.42\\ 
45 & 2.81 $\pm$ 1.01 & 0.90 $\pm$ 0.34 & 3.45 $\pm$ 0.97 & 3.30 $\pm$ 0.49 & 0.44 $\pm$ 0.50 & 0.91 $\pm$ 0.67\\ 
46 & 17.4 $\pm$ 0.79 & 6.54 $\pm$ 0.40 & 20.8 $\pm$ 1.01 & 20.3 $\pm$ 0.93 & 10.0 $\pm$ 0.83 & 21.9 $\pm$ 1.85\\ 
47 & 3.97 $\pm$ 1.37 & 1.66 $\pm$ 0.54 & 7.86 $\pm$ 2.36 & 6.19 $\pm$ 2.01 & 1.31 $\pm$ 0.90 & 1.94 $\pm$ 2.02\\ 
48 & 0.47 $\pm$ 0.73 & 0.05 $\pm$ 0.14 & 0.60 $\pm$ 1.57 & $...$ & 0.40 $\pm$ 0.96 & 1.09 $\pm$ 1.18\\ 
49 & 4.19 $\pm$ 0.39 & 1.83 $\pm$ 0.18 & 4.64 $\pm$ 0.74 & 4.81 $\pm$ 1.10 & 3.28 $\pm$ 1.40 & 1.51 $\pm$ 1.32\\ 
50 & 0.37 $\pm$ 7.50 & 0.37 $\pm$ 0.20 & 1.90 $\pm$ 1.51 & 2.01 $\pm$ 1.56 & 0.78 $\pm$ 1.17 & 0.91 $\pm$ 1.36\\ 
51 & 18.2 $\pm$ 1.36 & 6.27 $\pm$ 0.33 & 21.5 $\pm$ 1.00 & 19.3 $\pm$ 0.98 & 12.2 $\pm$ 0.58 & 6.54 $\pm$ 0.75\\ 
52 & 9.22 $\pm$ 0.80 & 2.83 $\pm$ 0.29 & 10.0 $\pm$ 1.26 & 9.88 $\pm$ 0.61 & 4.36 $\pm$ 0.63 & 5.59 $\pm$ 0.79\\ 
53 & 2.66 $\pm$ 1.12 & 0.97 $\pm$ 0.40 & 2.92 $\pm$ 1.56 & 3.62 $\pm$ 1.78 & 1.46 $\pm$ 1.49 & 1.33 $\pm$ 0.89\\ 
54 & 5.86 $\pm$ 0.88 & 2.16 $\pm$ 0.28 & 5.94 $\pm$ 0.79 & 6.65 $\pm$ 0.77 & 5.25 $\pm$ 1.13 & 1.19 $\pm$ 0.49\\ 
55 & 3.73 $\pm$ 0.56 & 1.08 $\pm$ 0.24 & 3.54 $\pm$ 1.00 & 5.52 $\pm$ 0.97 & 2.01 $\pm$ 0.55 & 2.86 $\pm$ 2.47\\ 
56 & 44.3 $\pm$ 5.43 & 8.94 $\pm$ 1.77 & 28.6 $\pm$ 5.48 & 29.0 $\pm$ 5.00 & 20.8 $\pm$ 3.42 & 16.9 $\pm$ 4.31\\ 
57 & 16.0 $\pm$ 4.62 & 4.44 $\pm$ 0.71 & 11.3 $\pm$ 2.86 & 8.66 $\pm$ 3.22 & 5.01 $\pm$ 1.34 & 1.37 $\pm$ 0.61\\ 
58 & 6.92 $\pm$ 0.60 & 2.20 $\pm$ 0.26 & 7.06 $\pm$ 0.87 & 8.03 $\pm$ 0.77 & 4.40 $\pm$ 0.49 & 3.69 $\pm$ 2.28\\ 
59 & 0.65 $\pm$ 0.57 & 0.24 $\pm$ 0.22 & 1.43 $\pm$ 0.99 & 0.32 $\pm$ 0.34 & 0.31 $\pm$ 0.34 & 1.48 $\pm$ 0.83\\ 
60 & 1.04 $\pm$ 0.51 & 0.35 $\pm$ 0.09 & 1.42 $\pm$ 0.49 & 1.04 $\pm$ 0.22 & 0.83 $\pm$ 0.72 & 2.13 $\pm$ 0.88\\ 
61 & 1.50 $\pm$ 0.43 & 0.60 $\pm$ 0.20 & 2.67 $\pm$ 0.79 & 1.11 $\pm$ 0.27 & 0.09 $\pm$ 0.47 & 0.41 $\pm$ 0.54\\ 
62 & 9.76 $\pm$ 1.21 & 3.45 $\pm$ 0.31 & 11.3 $\pm$ 0.95 & 13.0 $\pm$ 0.92 & 6.90 $\pm$ 0.49 & 4.58 $\pm$ 0.67\\ 
63 & 4.87 $\pm$ 0.66 & 1.67 $\pm$ 0.18 & 5.24 $\pm$ 0.62 & 4.39 $\pm$ 0.39 & 3.20 $\pm$ 0.49 & 1.85 $\pm$ 0.82\\ 
64 & 9.25 $\pm$ 1.76 & 3.14 $\pm$ 0.40 & 8.46 $\pm$ 1.31 & 11.5 $\pm$ 1.26 & 4.78 $\pm$ 1.26 & 4.92 $\pm$ 2.03\\ 
65 & 3.85 $\pm$ 0.56 & 1.14 $\pm$ 0.20 & 3.90 $\pm$ 1.09 & 3.05 $\pm$ 0.63 & 0.44 $\pm$ 0.16 & 2.16 $\pm$ 1.65\\
\enddata
\tablecomments{Integrated fluxes in W/m$^2$ from PAHFIT spectral decomposition for the PAH Emission features.  Values given as (...) do not have a measured flux from PAHFIT spectral decomposition.}
\end{deluxetable}

\begin{deluxetable}{ccccc}
\tablecaption{Fitted Fluxes of the Most Prominent Atomic Emission Lines\label{table line emission fits}}
\tablecolumns{5}
\tabletypesize{\footnotesize}
\tablewidth{0pc}
\tablehead{
\colhead{ID} & \colhead{F$\mathrm{_{[NeII]}}$} & \colhead{F$\mathrm{_{[NeIII]}}$} & \colhead{F$\mathrm{_{[OIV]}}$} & \colhead{F$\mathrm{_{[H_2 S(3)]}}$}\\ 
\colhead{} & \colhead{[$10^{-18}$]} & \colhead{[$10^{-18}$]} & \colhead{[$10^{-18}$]} & \colhead{[$10^{-18}$]}
}
\startdata
 1 & 8.84 $\pm$ 1.25 & 10.6 $\pm$ 1.30 & 1.52 $\pm$ 4.16 & 24.5 $\pm$ 20.9\\ 
 2 & $...$ & $...$ & 1.75 $\pm$ 0.49 & 3.15 $\pm$ 3.68\\ 
 3 & 2.25 $\pm$ 0.70 & $...$ & $...$ & 0.52 $\pm$ 1.23\\ 
 4 & 15.0 $\pm$ 3.55 & 3.56 $\pm$ 4.51 & 2.17 $\pm$ 0.59 & 9.78 $\pm$ 5.04\\ 
 5 & 11.2 $\pm$ 1.04 & 1.33 $\pm$ 3.02 & $...$ & 1.62 $\pm$ 1.57\\ 
 6 & 3.03 $\pm$ 0.76 & 0.17 $\pm$ 2.81 & 14.9 $\pm$ 5.49 & 0.02 $\pm$ 1.19\\ 
 7 & 26.0 $\pm$ 2.75 & 4.24 $\pm$ 2.33 & 7.54 $\pm$ 2.23 & 46.1 $\pm$ 9.36\\ 
 8 & 22.6 $\pm$ 2.07 & 12.0 $\pm$ 1.68 & $...$ & 0.48 $\pm$ 1.17\\ 
 9 & 14.5 $\pm$ 2.96 & 1.52 $\pm$ 3.35 & 1.70 $\pm$ 5.54 & 5.09 $\pm$ 4.81\\ 
10 & 14.0 $\pm$ 3.53 & 3.33 $\pm$ 4.20 & $...$ & 3.40 $\pm$ 4.24\\ 
11 & 7.65 $\pm$ 1.24 & $...$ & 1.25 $\pm$ 0.36 & 2.81 $\pm$ 1.66\\ 
12 & $...$ & $...$ & 0.81 $\pm$ 0.23 & 0.39 $\pm$ 0.70\\ 
13 & 25.2 $\pm$ 2.18 & 2.69 $\pm$ 2.36 & 5.97 $\pm$ 3.18 & 6.04 $\pm$ 3.00\\ 
14 & 6.46 $\pm$ 1.00 & 3.23 $\pm$ 3.01 & $...$ & 6.61 $\pm$ 5.54\\ 
15 & 5.74 $\pm$ 1.01 & $...$ & $...$ & 0.93 $\pm$ 1.85\\ 
16 & 2.49 $\pm$ 3.30 & 2.75 $\pm$ 5.95 & 23.2 $\pm$ 7.35 & 1.70 $\pm$ 3.21\\ 
17 & 3.37 $\pm$ 1.76 & 2.85 $\pm$ 2.14 & $...$ & 0.77 $\pm$ 4.28\\ 
18 & 17.7 $\pm$ 0.86 & 6.50 $\pm$ 0.91 & 2.24 $\pm$ 2.44 & 7.11 $\pm$ 2.98\\ 
19 & 1.05 $\pm$ 1.08 & 0.87 $\pm$ 1.64 & $...$ & 0.32 $\pm$ 2.08\\ 
20 & 51.3 $\pm$ 1.41 & 9.63 $\pm$ 1.94 & 2.30 $\pm$ 1.48 & 3.77 $\pm$ 1.19\\ 
21 & 16.3 $\pm$ 1.70 & 1.15 $\pm$ 4.65 & 1.04 $\pm$ 1.71 & 2.10 $\pm$ 2.06\\ 
22 &226.8 $\pm$ 1.73 & 11.5 $\pm$ 1.63 & 4.82 $\pm$ 16.5 & 8.08 $\pm$ 1.29\\ 
23 &235.8 $\pm$ 1.95 & 13.1 $\pm$ 1.68 & 11.6 $\pm$ 1.45 & 13.4 $\pm$ 3.34\\ 
24 & 1.33 $\pm$ 0.67 & 8.98 $\pm$ 1.50 & 6.43 $\pm$ 4.13 & 1.19 $\pm$ 1.76\\ 
25 & 25.8 $\pm$25.25 & 42.9 $\pm$25.34 & $...$ &25.99 $\pm$25.25\\ 
26 & 3.13 $\pm$ 1.46 & 0.89 $\pm$ 1.18 & 20.8 $\pm$ 0.67 & 0.62 $\pm$ 1.27\\ 
27 & 35.3 $\pm$ 1.67 & 8.80 $\pm$ 1.45 & 2.19 $\pm$ 1.61 & 7.71 $\pm$ 1.56\\ 
28 &28.11 $\pm$ 0.96 & 0.91 $\pm$ 0.77 & 2.60 $\pm$ 0.69 & 1.66 $\pm$ 1.51\\ 
29 & 9.12 $\pm$ 0.97 & 0.99 $\pm$ 0.91 & 1.04 $\pm$ 1.29 & 9.11 $\pm$ 2.84\\ 
30 & 27.4 $\pm$ 2.43 & 7.72 $\pm$ 2.65 & $...$ & 17.1 $\pm$ 3.86\\ 
31 & 2.65 $\pm$ 0.75 & 1.17 $\pm$ 2.78 & $...$ & 3.83 $\pm$ 5.45\\ 
32 & 9.22 $\pm$ 0.90 & 2.27 $\pm$ 0.70 & $...$ & 1.13 $\pm$ 1.74\\ 
33 & 1.99 $\pm$ 1.14 & 9.76 $\pm$ 40.0 & 3.06 $\pm$ 2.65 & 0.40 $\pm$ 1.32\\ 
34 & 0.58 $\pm$ 1.55 & 0.52 $\pm$ 1.40 & 0.52 $\pm$ 0.15 & 1.76 $\pm$ 1.98\\ 
35 & 16.7 $\pm$ 1.36 & 2.78 $\pm$ 1.51 & 1.68 $\pm$ 1.88 & 3.60 $\pm$ 2.64\\ 
36 & 3.35 $\pm$ 0.60 & 0.26 $\pm$ 0.87 & 1.91 $\pm$ 2.85 & 0.49 $\pm$ 3.05\\ 
37 & $...$ & 1.66 $\pm$ 1.51 & 0.51 $\pm$ 1.63 & $...$\\ 
38 & 86.2 $\pm$ 1.69 & 60.7 $\pm$ 3.56 & 12.4 $\pm$ 5.60 & 10.5 $\pm$ 1.40\\ 
39 & 1.99 $\pm$ 0.75 & 1.36 $\pm$ 1.01 & 2.52 $\pm$ 7.10 & 4.18 $\pm$ 4.63\\ 
40 & 6.87 $\pm$ 3.45 & 2.37 $\pm$ 3.43 & 3.17 $\pm$ 3.98 & 2.99 $\pm$ 6.70\\ 
41 & 6.07 $\pm$ 1.26 & 3.24 $\pm$ 1.87 & 0.50 $\pm$ 0.14 & 3.08 $\pm$ 0.99\\ 
42 & 42.4 $\pm$ 1.18 & 5.81 $\pm$ 0.94 & $...$ & 3.48 $\pm$ 2.45\\ 
43 & 24.3 $\pm$ 2.38 & 4.42 $\pm$ 2.69 & $...$ & 4.72 $\pm$ 2.74\\ 
44 & 0.65 $\pm$ 0.62 & 2.44 $\pm$ 0.74 & $...$ & 1.02 $\pm$ 1.92\\ 
45 & 6.30 $\pm$ 0.70 & $...$ & 2.90 $\pm$ 1.13 & 0.23 $\pm$ 1.42\\ 
46 & 20.1 $\pm$ 1.16 & 4.73 $\pm$ 1.29 & 1.28 $\pm$46.27 &46.71 $\pm$ 1.32\\ 
47 & 7.03 $\pm$ 1.30 & 2.10 $\pm$ 2.71 & 2.43 $\pm$ 5.68 & 14.0 $\pm$ 20.8\\ 
48 & 0.46 $\pm$ 4.35 & 0.51 $\pm$ 4.18 & 0.69 $\pm$ 0.19 & 13.3 $\pm$ 22.0\\ 
49 & 6.70 $\pm$ 1.90 & 2.56 $\pm$ 1.31 & 1.06 $\pm$ 0.29 & 1.95 $\pm$ 1.57\\ 
50 & 2.18 $\pm$ 1.50 & 0.96 $\pm$ 1.58 & 2.98 $\pm$ 1.86 & 2.49 $\pm$ 3.60\\ 
51 & 51.1 $\pm$ 0.80 & 7.58 $\pm$ 0.68 & 1.76 $\pm$ 1.31 & 4.08 $\pm$ 1.29\\ 
52 & 11.8 $\pm$ 0.98 & 6.27 $\pm$ 1.02 & 5.50 $\pm$ 1.14 & 5.18 $\pm$ 3.71\\ 
53 & 1.53 $\pm$ 1.79 & 6.88 $\pm$ 3.58 & 0.68 $\pm$ 0.19 & 1.29 $\pm$ 2.55\\ 
54 & 12.4 $\pm$ 1.51 & 11.6 $\pm$ 1.71 & 4.18 $\pm$ 1.33 & 1.11 $\pm$ 1.26\\ 
55 & 8.55 $\pm$ 1.32 & $...$ & 64.4 $\pm$ 18.3 & 1.82 $\pm$ 2.39\\ 
56 & 36.9 $\pm$ 3.98 & 9.21 $\pm$ 2.81 & $...$ & 10.2 $\pm$ 11.4\\ 
57 & 9.29 $\pm$ 1.65 & 6.66 $\pm$ 1.44 & $...$ & 17.3 $\pm$ 12.4\\ 
58 & 13.0 $\pm$ 0.87 & 4.23 $\pm$ 1.06 & 3.16 $\pm$ 2.89 & 0.94 $\pm$ 1.17\\ 
59 & $...$ & 0.36 $\pm$ 1.39 & 0.43 $\pm$ 0.12 & 1.18 $\pm$ 0.78\\ 
60 & 0.93 $\pm$ 0.62 & 0.61 $\pm$ 1.42 & 22.4 $\pm$ 5.32 & 0.42 $\pm$ 0.71\\ 
61 & 0.56 $\pm$ 0.55 & $...$ & 0.56 $\pm$ 0.15 & 0.19 $\pm$ 0.36\\ 
62 & 13.9 $\pm$ 0.61 & 2.05 $\pm$ 0.45 & 5.89 $\pm$ 1.07 & 3.40 $\pm$ 1.48\\ 
63 & 7.29 $\pm$ 0.65 & 1.84 $\pm$ 0.48 & 1.80 $\pm$ 1.05 & 1.16 $\pm$ 0.98\\ 
64 & 19.9 $\pm$ 1.60 & 2.96 $\pm$ 1.38 & $...$ & 6.93 $\pm$ 3.25\\ 
65 & 7.76 $\pm$ 0.93 & $...$ & 0.61 $\pm$ 0.17 & 3.18 $\pm$ 2.48\\
\enddata
\tablecomments{Integrated fluxes in W/m$^2$ from PAHFIT spectral decomposition for the atomic emission lines.  Values given as (...) do not have a measured flux from PAHFIT spectral decomposition.}
\end{deluxetable}

\begin{deluxetable}{ccccc}
\tablecaption{Equivalent Widths for Various PAH Emission Features}
\tablecolumns{5}
\tabletypesize{\footnotesize}
\tablewidth{0pc}
\tablehead{
\colhead{ID} & \colhead{EW$_{6.2}$} & \colhead{EW$_{7.7}$} & \colhead{EW$_{8.6}$} & \colhead{EW$_{11.3}$}
}
\startdata
 1 & 1.96 $\pm$ 0.22 & 4.46 $\pm$ 0.20 & 3.99 $\pm$ 0.11 & 0.07 $\pm$ 0.01\\ 
 2 & 0.02 $\pm$ 0.03 & 0.41 $\pm$ 0.12 & 0.05 $\pm$ 0.04 & 0.09 $\pm$ 0.02\\ 
 3 & 0.02 $\pm$ 0.07 & 0.60 $\pm$ 0.22 & 0.27 $\pm$ 0.05 & 0.10 $\pm$ 0.01\\ 
 4 & 0.36 $\pm$ 0.03 & 1.76 $\pm$ 0.11 & 0.76 $\pm$ 0.05 & 0.52 $\pm$ 0.04\\ 
 5 & 2.57 $\pm$ 1.67 & 7.03 $\pm$ 1.88 & 2.39 $\pm$ 0.36 & 2.54 $\pm$ 0.14\\ 
 6 & 0.56 $\pm$ 2.56 & 2.58 $\pm$ 1.51 & 0.64 $\pm$ 0.51 & 0.64 $\pm$ 0.09\\ 
 7 & 0.33 $\pm$ 0.02 & 2.23 $\pm$ 0.12 & 0.91 $\pm$ 0.05 & 1.42 $\pm$ 0.09\\ 
 8 & 1.81 $\pm$ 0.11 & 5.99 $\pm$ 0.30 & 1.65 $\pm$ 0.06 & 1.54 $\pm$ 0.03\\ 
 9 & 2.39 $\pm$ 1.30 & 6.09 $\pm$ 1.72 & 2.33 $\pm$ 0.46 & 2.34 $\pm$ 0.25\\ 
10 & 1.25 $\pm$ 0.48 & 3.63 $\pm$ 0.98 & 1.43 $\pm$ 0.30 & 2.23 $\pm$ 0.29\\ 
11 & 5.98 $\pm$ 0.65 & 9.76 $\pm$ 0.90 & 2.62 $\pm$ 0.20 & 2.11 $\pm$ 0.11\\ 
12 & 0.14 $\pm$ 0.03 & 1.09 $\pm$ 0.10 & 0.26 $\pm$ 0.04 & 0.17 $\pm$ 0.02\\ 
13 & 1.72 $\pm$ 0.14 & 6.84 $\pm$ 0.66 & 2.55 $\pm$ 0.17 & 2.93 $\pm$ 0.08\\ 
14 & 1.69 $\pm$ 0.73 & 4.08 $\pm$ 0.95 & 1.96 $\pm$ 0.36 & 4.03 $\pm$ 0.40\\ 
15 & 2.19 $\pm$ 4.17 & 6.45 $\pm$ 3.07 & 2.21 $\pm$ 0.50 & 2.37 $\pm$ 0.17\\ 
16 & 0.10 $\pm$ 0.05 & 0.47 $\pm$ 0.10 & 0.07 $\pm$ 0.04 & 0.20 $\pm$ 0.03\\ 
17 & 2.12 $\pm$ 1.68 & 2.63 $\pm$ 0.98 & 0.84 $\pm$ 0.31 & 0.53 $\pm$ 0.10\\ 
18 & 0.71 $\pm$ 0.06 & 1.72 $\pm$ 0.10 & 0.82 $\pm$ 0.04 & 0.98 $\pm$ 0.03\\ 
19 & 0.91 $\pm$ 64.86 & 4.04 $\pm$ 60.40 & 1.38 $\pm$ 5.30 & 1.68 $\pm$ 0.62\\ 
20 & 1.90 $\pm$ 0.20 & 5.43 $\pm$ 0.28 & 1.71 $\pm$ 0.04 & 1.75 $\pm$ 0.12\\ 
21 & 1.42 $\pm$ 0.23 & 5.73 $\pm$ 0.99 & 2.23 $\pm$ 0.29 & 2.67 $\pm$ 0.30\\ 
22 & 3.28 $\pm$ 0.16 & 10.27 $\pm$ 0.23 & 3.01 $\pm$ 0.04 & 1.91 $\pm$ 0.02\\ 
23 & 2.75 $\pm$ 0.07 & 10.80 $\pm$ 0.24 & 3.14 $\pm$ 0.05 & 2.55 $\pm$ 0.02\\ 
24 & 0.08 $\pm$ 0.03 & 0.68 $\pm$ 0.09 & 0.35 $\pm$ 0.04 & 0.30 $\pm$ 0.02\\ 
25 & 2.10 $\pm$ 0.27 & 6.26 $\pm$ 0.37 &25.87 $\pm$ 0.06 &25.86 $\pm$ 0.03\\ 
26 & 0.65 $\pm$ 0.15 & 2.60 $\pm$ 0.27 & 0.85 $\pm$ 0.07 & 0.84 $\pm$ 0.04\\ 
27 & 1.91 $\pm$ 0.11 & 7.41 $\pm$ 0.24 & 2.40 $\pm$ 0.05 & 1.51 $\pm$ 0.03\\ 
28 & 0.93 $\pm$ 0.06 & 3.64 $\pm$ 0.31 & 1.41 $\pm$ 0.08 & 2.35 $\pm$ 0.10\\ 
29 & 1.23 $\pm$ 0.09 & 5.30 $\pm$ 0.36 & 2.02 $\pm$ 0.13 & 1.92 $\pm$ 0.12\\ 
30 & 0.61 $\pm$ 0.03 & 2.51 $\pm$ 0.15 & 1.20 $\pm$ 0.06 & 1.86 $\pm$ 0.06\\ 
31 & 0.35 $\pm$ 0.66 & 1.04 $\pm$ 0.90 & 0.51 $\pm$ 0.27 & 1.33 $\pm$ 0.14\\ 
32 & 0.81 $\pm$ 0.36 & 2.10 $\pm$ 0.51 & 0.97 $\pm$ 0.12 & 1.76 $\pm$ 0.09\\ 
33 & 0.13 $\pm$ 0.03 & 0.58 $\pm$ 0.08 & 0.24 $\pm$ 0.04 & 0.20 $\pm$ 0.02\\ 
34 & 0.32 $\pm$ 1.20 & 2.68 $\pm$ 1.06 & 1.29 $\pm$ 0.30 & 0.21 $\pm$ 0.06\\ 
35 & 0.90 $\pm$ 0.07 & 3.18 $\pm$ 0.24 & 1.20 $\pm$ 0.07 & 1.74 $\pm$ 0.05\\ 
36 & 0.35 $\pm$ 0.36 & 1.41 $\pm$ 0.61 & 0.48 $\pm$ 0.17 & 0.77 $\pm$ 0.10\\ 
37 & 2.50 $\pm$ 0.63 & 6.61 $\pm$ 1.26 & 1.59 $\pm$ 0.31 & 1.69 $\pm$ 0.24\\ 
38 & 1.89 $\pm$ 0.06 & 7.01 $\pm$ 0.16 & 1.88 $\pm$ 0.03 &38.14 $\pm$ 0.01\\ 
39 & 0.35 $\pm$ 2.53 & 2.00 $\pm$ 1.93 & 0.95 $\pm$ 0.54 & 0.84 $\pm$ 0.13\\ 
40 & 6.17 $\pm$ 7.58 & 8.36 $\pm$ 4.45 & 2.21 $\pm$ 0.88 & 1.33 $\pm$ 0.23\\ 
41 & 0.72 $\pm$ 0.32 & 2.68 $\pm$ 0.56 & 1.01 $\pm$ 0.18 & 3.20 $\pm$ 0.25\\ 
42 & 2.93 $\pm$ 0.61 & 7.50 $\pm$ 0.97 & 2.29 $\pm$ 0.22 & 1.12 $\pm$ 0.14\\ 
43 & 1.17 $\pm$ 0.11 & 4.05 $\pm$ 0.30 & 1.13 $\pm$ 0.06 & 0.86 $\pm$ 0.05\\ 
44 & 0.07 $\pm$ 0.06 & 0.64 $\pm$ 0.30 & 0.05 $\pm$ 0.13 & 0.25 $\pm$ 0.02\\ 
45 & 0.61 $\pm$ 0.26 & 1.84 $\pm$ 0.39 & 0.71 $\pm$ 0.10 & 0.88 $\pm$ 0.05\\ 
46 & 2.04 $\pm$ 0.33 & 7.26 $\pm$ 0.67 & 2.32 $\pm$ 0.13 & 1.97 $\pm$ 0.14\\ 
47 & 1.73 $\pm$ 10.84 & 5.96 $\pm$ 6.75 & 4.64 $\pm$ 2.36 & 4.56 $\pm$ 0.69\\ 
48 & 0.25 $\pm$ 1.00 & 0.19 $\pm$ 0.47 & 0.21 $\pm$ 0.26 & 0.01 $\pm$ 0.06\\ 
49 & 2.39 $\pm$ 2.13 & 13.13 $\pm$ 2.43 & 3.25 $\pm$ 0.43 & 2.06 $\pm$ 0.22\\ 
50 & 0.19 $\pm$ 10.89 & 1.74 $\pm$ 27.35 & 0.97 $\pm$ 4.52 & 1.30 $\pm$ 0.45\\ 
51 & 2.71 $\pm$ 0.51 & 7.61 $\pm$ 0.39 & 2.52 $\pm$ 0.09 & 2.30 $\pm$ 0.04\\ 
52 & 0.91 $\pm$ 0.08 & 2.76 $\pm$ 0.38 & 1.27 $\pm$ 0.11 & 2.18 $\pm$ 0.17\\ 
53 & 1.00 $\pm$ 0.37 & 2.97 $\pm$ 0.58 & 0.90 $\pm$ 0.15 & 1.20 $\pm$ 0.10\\ 
54 & 1.26 $\pm$ 0.10 & 4.08 $\pm$ 0.25 & 1.09 $\pm$ 0.06 & 1.09 $\pm$ 0.04\\ 
55 & 2.35 $\pm$ 2.39 & 4.32 $\pm$ 1.61 & 1.50 $\pm$ 0.33 & 2.54 $\pm$ 0.26\\ 
56 & 2.04 $\pm$ 0.49 & 4.47 $\pm$ 0.68 & 1.85 $\pm$ 0.21 & 2.35 $\pm$ 0.14\\ 
57 & 1.65 $\pm$ 0.35 & 3.91 $\pm$ 0.52 & 1.57 $\pm$ 0.57 & 0.98 $\pm$ 0.10\\ 
58 & 4.12 $\pm$ 1.17 & 8.85 $\pm$ 1.27 & 2.64 $\pm$ 0.27 & 2.97 $\pm$ 0.15\\ 
59 & 0.20 $\pm$ 0.16 & 0.85 $\pm$ 0.18 & 0.54 $\pm$ 0.09 & 0.15 $\pm$ 0.03\\ 
60 & 0.21 $\pm$ 0.09 & 0.71 $\pm$ 0.15 & 0.33 $\pm$ 0.06 & 0.39 $\pm$ 0.03\\ 
61 & 0.37 $\pm$ 0.06 & 1.92 $\pm$ 0.15 & 0.95 $\pm$ 0.07 & 0.49 $\pm$ 0.03\\ 
62 & 1.09 $\pm$ 0.12 & 4.02 $\pm$ 0.42 & 1.47 $\pm$ 0.10 & 2.26 $\pm$ 0.06\\ 
63 & 3.76 $\pm$ 3.92 & 11.72 $\pm$ 2.02 & 2.92 $\pm$ 0.28 & 1.22 $\pm$ 0.04\\ 
64 & 0.89 $\pm$ 0.11 & 2.79 $\pm$ 0.25 & 0.92 $\pm$ 0.06 & 1.41 $\pm$ 0.04\\ 
65 & 4.75 $\pm$ 1.04 & 6.75 $\pm$ 1.18 & 2.09 $\pm$ 0.29 & 1.53 $\pm$ 0.10\\
\enddata
\tablecomments{Equivalent widths are from PAHFIT spectral decomposition results and the errors were done using monte carlo simulations and running through PAHFIT to get uncertainties.  Values given as (...) do not have a measured flux from PAHFIT spectral decomposition.}
\end{deluxetable}

\clearpage
\appendix
\section{Estimating the Total IR luminosity}
\label{MAGES LIR}
At the redshifts of our sample, $0.02 <$ z $< 0.6$, the MIPS IR data probe the rest-frame mid-to-far-IR, which broadly correlates with the total IR luminosity \citep[e.g.][]{CharyElbaz2001, Papovich2002, Rieke2009}.  We used models \citep{CharyElbaz2001, DaleHelou2002, Rieke2009} for the IR SED of galaxies to convert the observed MIPS data to total IR luminosity.  By selection, all galaxies in our sample have 24\micron\ detections, and we convert the observed 24\micron\ flux densities to rest-frame luminosity densities at $24 \micron /(1+z)$.  We then correct these values to a total IR luminosity, \lir(24\micron), using \textit{each} of the three sets of model IR SEDs, assuming for each set of models that a given rest-frame $24 \micron /(1+z)$ luminosity density translates uniquely to a single SED template.  The \citet{Rieke2009} templates apply to local galaxies and do not include the SED evolution seen at high redshift.  However, \citet{Rujopakarn2012} show that the local SEDs and those appropriate for high redshift converge for the redshift range of our sample.

For those galaxies with detections at 70\micron\ and/or 160\micron\ in the MAGES data (as listed in table \ref{table galaxy stats}), we use all available MIPS flux densities to estimate the total IR luminosity.  For each source, we convert the observed MIPS data to rest-frame luminosity densities, and scaled each IR SED in each set of models with a multiplicative factor to obtain a best-fit for each SED.  For each set of models, we minimized a  $\chi^2$ statistic with $\chi_a^2 = \sum ( d_i - c \times m^a_i)^2 / \sigma_i$, where $d_i$ is the rest-frame luminosity density measured from the $i$=24\micron, 70\micron, and 160\micron\ data, $m_{(a,i)}$ is the luminosity density expected for model IR SED $a$, $c$ is the fitted scale factor, and $\sigma_i$ are the errors on the measured MIPS data.  We chose the IR SED from each set of models that minimizes $\chi^2_a$.

\begin{figure}
\epsscale{1.15}
\plottwo{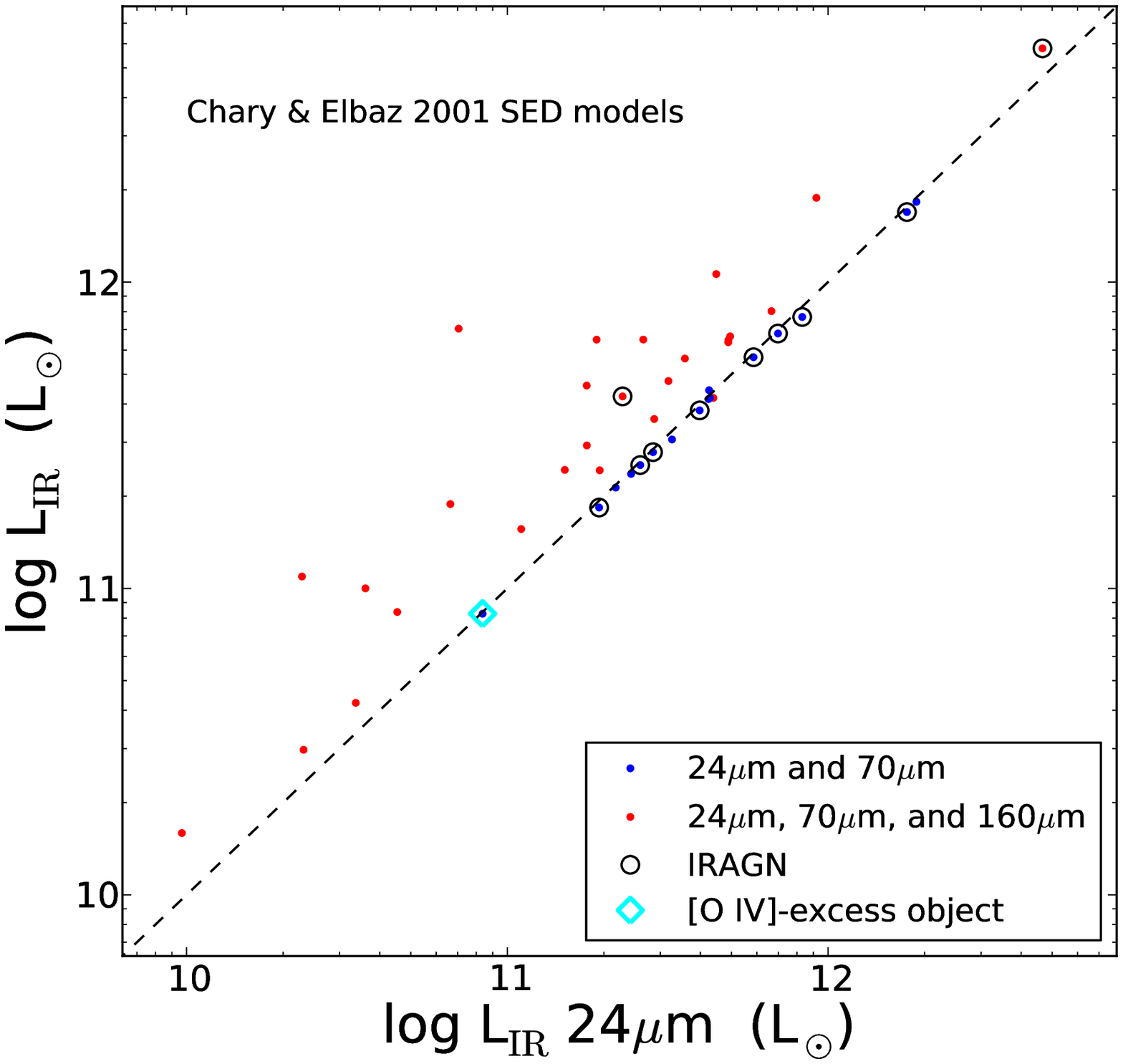}{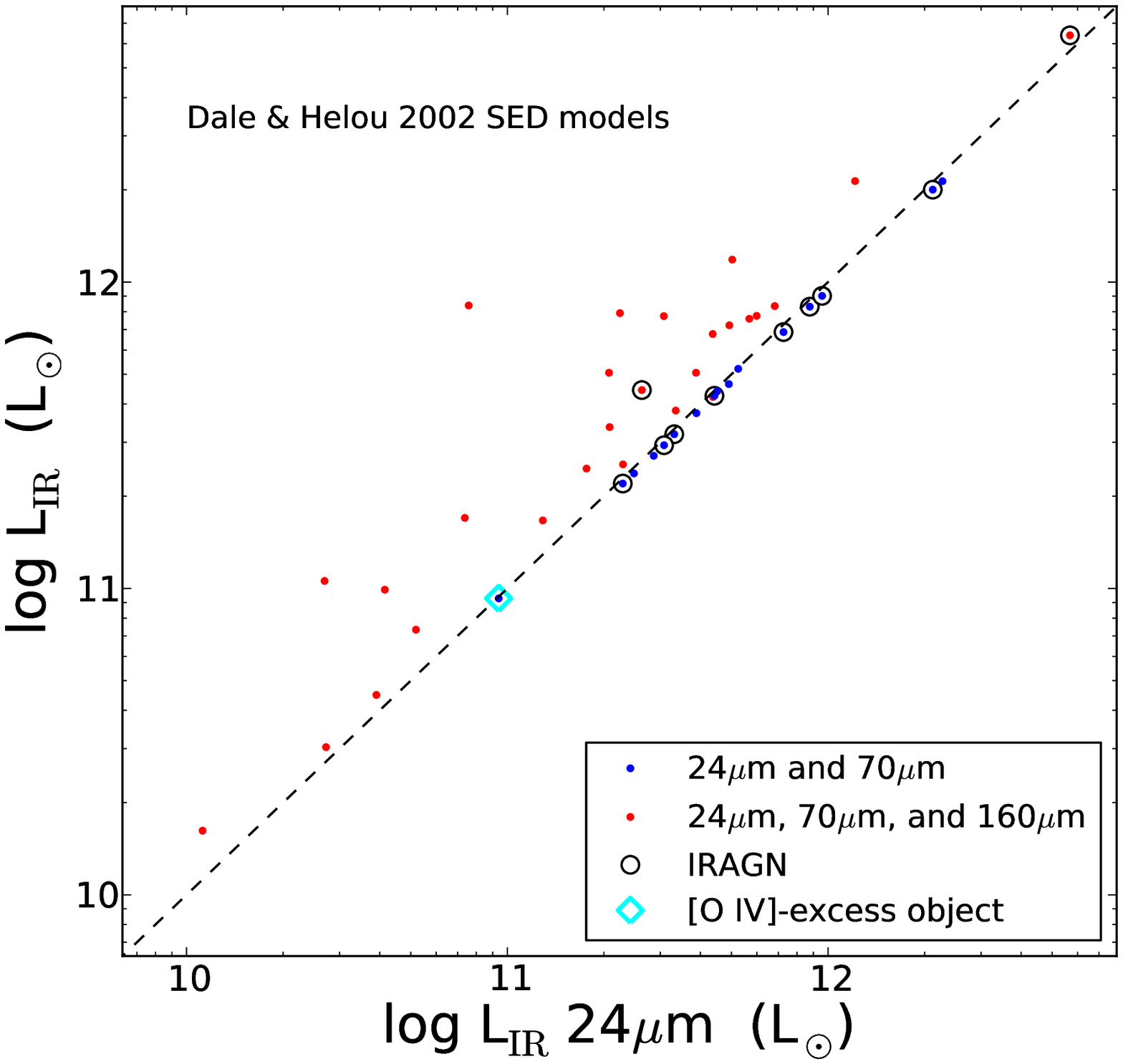}
\epsscale{0.6}
\plotone{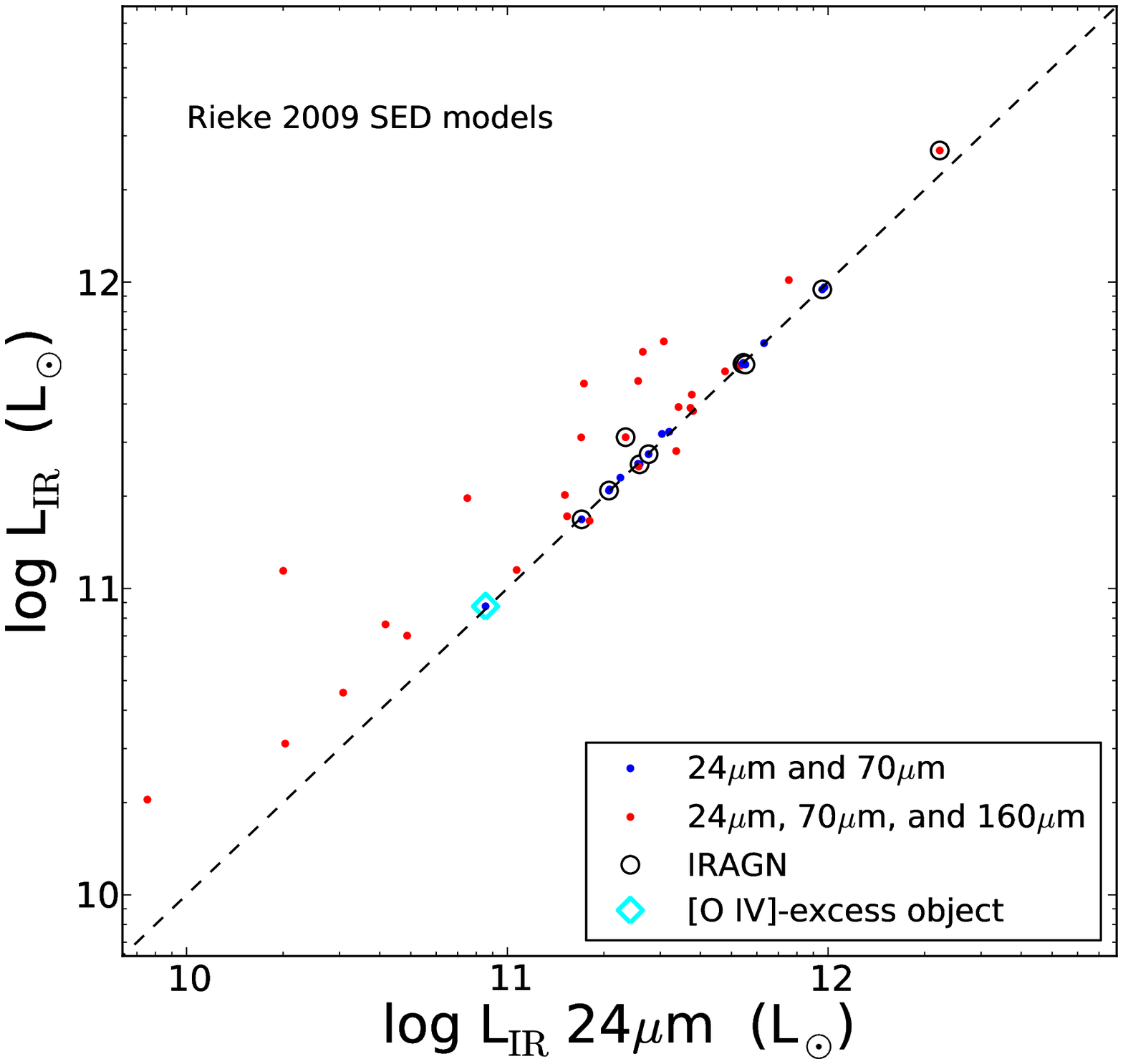}
\caption{ The IR luminosities derived from the 24\micron\ data only \lir(24\micron) compared to the \lir\ derived using multiple MIPS bands (24\micron, 70\micron\ and/or 160\micron) for each of the three sets of IR SED models \citep[][as labeled in each panel]{CharyElbaz2001, DaleHelou2002, Rieke2009}.  The dashed line represents a unity relation between the two \lir\ estimates.  Sources with only 24\micron\ flux densities are not plotted.}
\label{LIR all}
\end{figure}

In figure \ref{LIR all}, we compare the \lir(24\micron) against the \lir\ derived using multiple MIPS bands (24\micron, 70\micron\ and/or 160\micron)  for each of the three sets of IR SED models \citep{CharyElbaz2001, DaleHelou2002, Rieke2009}.  We found that there is no more than a factor of 2-3 difference in the derived IR luminosity from multiple MIPS bands and that derived from 24\micron\ only between any of the IR SED models.  This is particularly true for galaxies with total IR luminosities \lir\ $> 10^{11}$\lsol\ (the majority of \lir\ estimates in our sample).  However, our far-IR (160\micron\ sources) sample is incomplete in the direction that the less luminous galaxies (\lir\ $< 10^{11}$\lsol) will be under-represented.  As illustrated in figure~\ref{LIR all} and discussed below, the IR luminosities derived from the 24\micron\ data only from the \citet{Rieke2009} SED templates are the most consistent with the IR luminosities derived using multiple MIPS bands.  In particular, for IRAGN the IR luminosities derived with the \citet{Rieke2009} SED templates with only 24\micron\ data are consistent with those from multiple MIPS bands.  We find the median ratio estimates of the total IR luminosity (using all bands available) compared to the total IR luminosity (using only the 24\micron\ band) for the \citet{CharyElbaz2001}, \citet{DaleHelou2002}, and \citet{Rieke2009} models to be 0.97, 0.96, and 1.0, respectively.

We considered the idea that the IR SED templates we used for the IRAGN may be incorrect, which could be the case for those IRAGN with only 24\micron\ detections.   Specifically, we computed the total IR luminosities from the 24 micron flux densities for the IRAGN using the SEDs for optically luminous quasars \citep{Shang2011}.  In these cases the total IR luminosities for the IRAGN would decrease by at most a factor of five (at most a factor of 3.5 for the IRAGN with only 24\micron\ detections).   Because there are only four IRAGN with only 24\micron\ detections, and the effect is not large ($\approx$0.5 dex in \lir), this would not change any of our conclusions if all of these IRAGN have SEDs of optical luminous quasars.  However, the MIPS-to-IRAC colors of the IRAGN in our sample (F$_{24\micron}$/F$_{8.0\micron} = 4.2$ median and ranging from 3.5 to 6.1 for the four IRAGN with only 24\micron\ detections) are highly discrepant compared to those of the \citet{Shang2011} optically luminous SEDs (F$_{24\micron}$/F$_{8.0\micron} = 1.4$ median).  Therefore, it seems highly unlikely that the IRAGN in our sample have SEDs like optical QSOs\footnote{Indeed, we have an optical spectrum for one of the IRAGN with only 24\micron\ detection and it only shows narrow lines indicative of a Seyfert galaxy, clearly not a broad-line quasar.}.  In this case, we take the lower IR luminosity of a factor of 3.5 to be a lower bound for an extreme case for the IRAGN in our sample.

We found that when only the 24\micron\ and 70\micron\ flux densities are available the estimates of the total IR luminosity are similar to those derived using 24\micron\ only for the \citet{Rieke2009} SED models.  And when using the \citet{CharyElbaz2001} and \citet{DaleHelou2002} models the estimates for the total IR luminosity are smaller compared to the that derived with only the 24\micron\ flux density by a median of 0.97 and 0.95, respectively.  When all three MIPS flux densities are available the estimate of the total IR luminosity is larger compared to that derived by 24\micron\ only (larger by a median of 1.34, 1.61, and 1.44 for the \citet{Rieke2009}, \citet{CharyElbaz2001}, and \citet{DaleHelou2002} SED models, respectively).  One conclusion from this analysis is that objects in our sample detected at 160\micron\ appear to have a large cold-dust mass which is not sufficiently probed by the 24\micron\ and 70\micron\ bandpasses.  However, since the 160\micron\ detections are not complete and preferentially find galaxies with larger-than-typical flux densities, the larger luminosity estimates may have a contribution due to selection bias.

For these reasons, we adopt the IR luminosities derived using the \citet{Rieke2009} SED templates for this work.  For all galaxies in our sample, we use the IR luminosity derived from all available MIPS bands.

\end{document}